\documentclass[%
 reprint,
 amsmath,amssymb,
 aps,
]{revtex4-2}
\usepackage{tikz}
\usetikzlibrary{quantikz2}
\usepackage[dvipsnames]{xcolor}
\definecolor{mygreen}{HTML}{00A64F}
\definecolor{mymagenta}{HTML}{EC008C}
\definecolor{mycyan}{HTML}{00AEEF}
\tikzset{%
    uctrl/.style={draw, circle, minimum size=0.285pc, append after command={ \pgfextra { \fill  (0,0)-- (225:2pt) arc (225:405:2pt) -- cycle; } } }
}

\DeclareExpandableDocumentCommand{\uctrl}{O{}{m}}{|[uctrl,#1]| {#2} \qw}
\usepackage{dsfont}
\usepackage{adjustbox}
\usepackage{braket}
\usepackage{babel}
\usepackage{physics}
\usepackage{tabularx}
\newcolumntype{Y}{>{\centering\arraybackslash}X}
\usepackage{siunitx}
\usepackage[utf8]{inputenc}
\usepackage{nicematrix}
\usepackage{graphicx}
\usepackage{dcolumn}
\usepackage{bm}
\usepackage{booktabs}
\usepackage{float}
\usepackage{hyperref}

\usepackage{lipsum, babel}

\usepackage{todonotes}

\begin{document}

\title{End-to-End Quantum Algorithm for Topology Optimization in Structural Mechanics}

\author{Leonhard H\"olscher$^{1,2,3}$}
\email{leo.h@mail.de}
\author{Oliver Ahrend$^{4}$}
\author{Lukas Karch$^{1}$}
\author{Carlotta L'Estocq$^{1}$}
\author{Marc Marfany Andreu$^{1}$}
\author{Tobias Stollenwerk$^{3}$}
\author{Frank K. Wilhelm$^{2,3}$}
\author{Julia Kowalski$^{4}$}

\affiliation{$^{1}$BMW Group, 80809 Munich, Germany}
\affiliation{$^{2}$Theoretical Physics, Saarland University, 66123 Saarbr\"ucken, Germany}
\affiliation{$^{3}$Institute for Quantum Computing Analytics (PGI 12), Forschungszentrum J\"ulich, 52425 J\"ulich, Germany}
\affiliation{$^{4}$ Chair of Methods for Model-based Development in Computational Engineering, RWTH Aachen University, 52062 Aachen, Germany}
\date{\today}

\begin{abstract}
Topology optimization is a key methodology in engineering design for finding efficient and robust structures. Due to the enormous size of the design space, evaluating all possible configurations is typically infeasible. In this work, we present an end-to-end, fault-tolerant quantum algorithm for topology optimization that operates on an exponentially large Hilbert space representing the design space. We demonstrate the algorithm on the two-dimensional Messerschmitt-Bölkow-Blohm (MBB) beam problem. By restricting design variables to binary values, we reformulate the compliance minimization task as a combinatorial satisfiability problem, solved using Grover's algorithm. Within Grover's oracle, the compliance is computed through the finite-element method (FEM) using established quantum algorithms, including block-encoding of the stiffness matrix, Quantum Singular Value Transformation (QSVT) for matrix inversion, Hadamard test, and Quantum Amplitude Estimation (QAE). The complete algorithm is implemented and validated using classical quantum-circuit simulations. A detailed complexity analysis shows that the method evaluates the compliance of exponentially many structures in quantum superposition in polynomial time. In the global search, our approach maintains Grover's quadratic speedup compared to classical unstructured search. Overall, the proposed quantum workflow demonstrates how quantum algorithms can advance the field of computational science and engineering.
\end{abstract}

\maketitle
\section{Introduction}\label{sec:intro}
Topology optimization (TO) is a computational design methodology that determines the optimal material distribution within a design space to maximize structural performance of the system under given load and boundary conditions and possibly further constraints \cite{bendsoe_topology_2004, sigmund_topology_2013}. 
Structural performance can be measured in various ways and depends on the exact task, for example structural stiffness against deformation, minimizing mass or material volume, or excited and avoided eigenfrequencies.
By integrating simulation-based performance and optimization, TO enables the creation of lightweight and robust structures across engineering applications, including aerospace, automotive, and additive manufacturing. 

In its standard formulation, TO discretizes the design space into a so called conforming mesh, i.e. the cells are non-overlapping and there are no gaps between adjacent cells. The objective is then to find the optimal configuration of cells that should contain material. This optimization problem is typically formalized in either of two ways. The first using a binary decision variable for each cell that denotes whether the cell contains material or not. This might also be extended to multiple materials, where the binary variable is replaced by an integer variable, denoting the material or void. The second approach uses a continuous variable to represent the extent to which the cell should be filled, sometimes referred to as density - not to be confused with the physical density of the actual material. Then, the Finite-Element-Method (FEM) is employed to evaluate the system performance of a candidate material distribution. Without loss of generality, TO can also be applied to other fields, for example fluid mechanics and electromagnetism, using other modeling frameworks, for example Computational-Fluid-Dynamics (CFD) or laboratory experiments. As our use-case originates from structural mechanics, we restrict our paper to FEM-based TO. Sigmund \& Maute \cite{sigmund_topology_2013} give a review of the foundations of TO, and discuss different methods and implications.\\


Despite its widespread success, TO remains computationally demanding. Each candidate design's structural performance is evaluated by solving a large system of linear equations, and the number of possible candidates grows exponentially with the number of cells of the dicretized design space. To manage this complexity, most classical implementations make use of the continuous approach using cell densities between zero and one. This allows the use of gradient-based solvers but introduces intermediate, non-interpretable material states, where the cell densities must later be rounded to zero or one, or penalized right away during the optimization \cite{sigmund_99_2001, andreassen_efficient_2011}. This is because a cell being partially filled with material imposes another TO problem to find the respective material distribution within the cell. In contrast, the binary formulation preserves physical interpretability but turns the optimization into an exponentially hard combinatorial problem. Evaluating vast numbers of structural configurations, each involving a costly FEM simulation, represents a central computational bottleneck.

Quantum computing offers a fundamentally different route to overcome this challenge. By encoding each candidate structure as a quantum state, a quantum computer can exploit an exponentially large Hilbert space to process all possible structures in superposition. Grover's algorithm \cite{grover_fast_1996}, for example, naturally exploits this exponential space for global search, enabling a quadratic speedup in identifying optimal or near-optimal configurations. At the same time, repeatedly solving large systems of linear equations arising from the FEM analysis is another major computational bottleneck in TO. Here, quantum linear system algorithms, such as the Harrow-Hassidim-Lloyd (HHL) algorithm \cite{harrow_quantum_2009}, the Quantum Singular Value Transform (QSVT) \cite{gilyen_quantum_2019, martyn_grand_2021}, and others \cite{costa_optimal_2022, liu_improved_2025, dalzell_shortcut_2024} have the potential to achieve exponential speedups. Combining the combinatorial search over candidate structures and a simulation-based performance assessment of the underlying physical system could benefit from substantial quantum acceleration. However, the central challenge is to integrate these components into a coherent quantum workflow without introducing (classical) overheads that would eliminate any achievable speedup \cite{aaronson_read_2015}, for example, state preparation of large matrices or read-outs of displacement vectors.

Recent studies have started to investigate the integration of quantum algorithms with TO.
Sato et al. \cite{sato_quantum_2023} successfully demonstrated the application of Variational-Quantum-Algorithms (VQAs) to solve the binary TO setting. The variational approach shifts the complexity of solving a combinatorial problem to a continuous problem in the parameters of quantum gates, i.e. rotation angles. While this approach can be tuned to provably provide the best solution of the initial combinatorial problem \cite{bravo-prieto_variational_2023}, in practice, the continuous optimization problem is often still intractable due to its highly nonlinear nature. Specialized classical optimizers are being developed to tackle this problem \cite{iannelli_noisy_2024}, but an overall quantum advantage remains questionable. 
Ye and Pan \cite{ye_towards_2025} reformulated TO with multiple materials as a mixed-integer linear program solved via a Modified Dantzig-Wolfe decomposition \cite{dantzig_decomposition_1960}, where local binary sub-problems are expressed as Quadratic-Unconstrained-Binary-Optimization problems (QUBOs) for quantum annealing. 
Sukulthanasorn et al. \cite{sukulthanasorn_novel_2025} propose to reduce the number of equation systems required to solve by leveraging an updater, which is tasked to find the optimal distribution of material with as few evaluations of the system performance as possible. The updater is formulated as a QUBO problem, which is then solved by quantum annealing. 
Beyond quantum approaches specifically targeting TO, Stein et al. \cite{stein_exponential_2023} proposed Quantum Simulation-based Optimization (QuSO), embedding numerical simulations or the corresponding system of linear equations within the Quantum Approximate Optimization Algorithm (QAOA). The QuSO algorithm has been successfully applied to optimize power grids \cite{adler_scaling_2025} and automotive cooling systems \cite{holscher_quantum_2025}, highlighting the potential and limitations of this framework. However, the variational nature of QAOA does not allow us to give a general statement about any quantum advantage.

In this work, we present a fault-tolerant quantum algorithm for TO that combines FEM analysis and global optimization within a single coherent algorithm. By fault-tolerant, we refer to the class of quantum algorithms meant to be executed on error-corrected quantum computers. Using the well-known bending beam problem, also called the Messerschmitt-Bölkow-Blohm (MBB) beam problem \cite{sigmund_99_2001, andreassen_efficient_2011, sigmund_topology_2013}, as a demonstrator, we reformulate TO as a quantum search over binary material configurations. Each candidate design is represented as a quantum state, and Grover's algorithm is employed to identify optimal configurations. Within Grover's oracle, the structural performance is quantified through the compliance, which is computed using a block-encoded stiffness matrix, QSVT for matrix inversion, the Hadamard test, and Quantum Amplitude Estimation (QAE). In contrast to previous quantum approaches to TO that rely on quantum annealing or VQAs, our method is composed entirely of fault-tolerant quantum subroutines. This allows us to derive a full algorithmic complexity analysis demonstrating that the overall workflow maintains Grover's quadratic speedup compared to classical unstructured search.

The remainder of this paper is organized as follows. Section \ref{sec:toopt} reviews the classical TO framework and its FEM formulation, including the discretization of the design space and the binary and density-based representation of the material distribution. This motivates the binary formulation adopted for our quantum implementation. Section \ref{sec:quantalg} introduces the proposed quantum algorithm and explains each component of the workflow, from problem definition to circuit-level realization. Section \ref{sec:numerics} presents numerical experiments that demonstrate the algorithm's functionality. Section \ref{sec:discussion} discusses the implications, limitations, and possible extensions of the approach. Finally, Section \ref{sec:outlook} summarizes the main findings and outlines directions for future research.

\section{Topology Optimization}\label{sec:toopt}
The objective in this TO is to design structures that deform as little as possible under given supports and loads. A common performance measure is the compliance \cite{bendsoe_topology_2004, sigmund_topology_2013}
\begin{equation}
    c(\rho) = \int_\Omega \mathbf{u}(\rho, \mathbf{r})\cdot\boldsymbol{f}(\mathbf{r}) \,\mathrm{d}\mathbf{r},
    \label{eq:compliance_continuous}
\end{equation}
where $\rho(\mathbf{r}) \in [0,1]$ denotes the relative density of material, as a function of spatial coordinates $\mathbf{r}$, $\mathbf{u}(\rho, \mathbf{r})$ the displacement field, and $\boldsymbol{f}(\mathbf{r})$ the external load over the domain $\Omega$. Eq.\,\eqref{eq:compliance_continuous} shows, that the compliance measures the deformation in the same direction as the load. Deformation orthogonal to the load does not contribute to the compliance. As a sidenote, minimizing compliance is the same as maximizing stiffness. To avoid the trivial all-solid design, one typically constrains the total amount of material. We want to emphasize that relative density $\rho$ is independent of the material density and is introduced only for the definition of the minimization problem. Therefore, without loss of generality $\rho$ may be dimensionless, and the integral
\begin{equation}
    V(\rho) = \int_\Omega \rho \,\mathrm{d}\mathbf{r}\bigg/ \int_\Omega 1 \,\mathrm{d}\mathbf{r}
\end{equation}
is interpreted as the fraction of volume containing material. 
The continuous optimization problem then reads
\begin{equation}
\begin{split}
    \min_\rho \ & c(\rho), \\
    \text{subject to } & V(\rho) = V_0,
\end{split}
\end{equation}
where $V_0\in(0,1)$ specifies the target volume fraction.

\subsection{Finite Element Discretization}
\begin{figure}[h]
    \centering
    \includegraphics{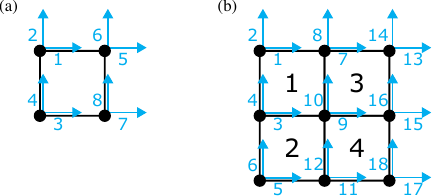}
    \caption{FEM domain discretization using quadrilateral elements. (a) shows a single element with four nodes, each having a horizontal and a vertical degree of freedom (DoF), and (b) depicts a domain consisting of $2\times2$ elements. Either way, we order the DoFs by counting column-wise from top to bottom from left to right. Horizontal DoFs have precedence over vertical DoFs.}
    \label{fig:element}
\end{figure}

In this paper, we consider the 2D MBB beam problem, which is a common benchmark in TO \cite{sigmund_99_2001, andreassen_efficient_2011, sigmund_topology_2013}. We discretize a two-dimensional beam into $n_x\times n_y$ identical cells, where $n_x$ and $n_y$ is the number of cells in horizontal $x$- and vertical $y$-direction, respectively. For the sake of simplicity, each cell is then used as an element for the FEM. This is not a requirement of the TO, and one could remesh the cells for an arbitrary FEM discretization. However, maintaining a fully structured, equidistant, FEM mesh significantly simplifies the matrix encoding in the quantum algorithm. Therefore, we use bilinear quadrilateral elements that have four nodes each with two translational degrees of freedom (DoFs) per node. Fig.\,\ref{fig:element} illustrates the FEM discretization. 
Thus, the element force vector $\mathbf{f}^\text{el}$ and the resulting element displacement vector $\mathbf{u}^\text{el}$ have eight entries, ordered according to Fig.\,\ref{fig:element}. In the linear elasticity regime, force and displacement are related linearly by 
\begin{equation}
    \mathbf{K}^\text{el}\mathbf{u}^\text{el}=\mathbf{f}^\text{el},
    \label{eq:Keuefe}
\end{equation}
where $\mathbf{K}^\text{el}$ is the $8\times8$ element stiffness matrix. Entry $\mathbf{K}^\text{el}_{i,j}$ relates the displacement of DoF $j$ $(\mathbf{u}^\text{el}_j)$ to the force acting in direction of DoF $i$ $(\mathbf{f}^\text{el}_i)$, and physically has units of force per length. More details on $\mathbf{K}^\text{el}$ are given in the Appendix\,\ref{appendix:derivationstiffness}. 

The domain is assembled by stitching the $n_\text{el}=n_x n_y$ elements together. Since neighboring elements share two nodes, we need to introduce a global node ordering. Therefore, we count column-wise from top to bottom, then left to right, as shown in Fig.\,\ref{fig:element}(b). This gives us the global displacement and force vector $\mathbf{u}, \mathbf{f}\in\mathbb{R}^{n_\text{DoF}}$ with the total number of DoFs 
\begin{equation}
    n_\text{DoF}=2(n_x+1)(n_y+1).
\end{equation}
Similarly to Eq.\,\eqref{eq:Keuefe}, we have a global linear system of equations
\begin{equation}
    \mathbf{K}\mathbf{u}=\mathbf{f},
    \label{eq:Kuf}
\end{equation}
where the global stiffness matrix $\mathbf{K}$ is obtained by mapping element stiffness matrices into the global DoF space and summing all element contributions. This is also referred to as assembling the global stiffness matrix. In the context of TO, stiffness matrices depend on the material distribution, introduced as $\rho$ above, as outlined in the following.

\subsection{Density-Based Formulation}
In common density-based approaches, such as the Solid Isotropic Material with Penalization (SIMP) method \cite{andreassen_efficient_2011}, the material distribution is represented by ${\mathbf{x}=(x_1,\dots,x_{n_\text{el}})^\top}$ with $x_e \in [0,1]$ denoting the relative density of element $e$. Here and in the remainder, $\mathbf{x}$ is the discretized version of $\rho$.
The global stiffness matrix then reads
\begin{equation}
    \mathbf{K}(\mathbf{x}) = \sum_{e=1}^{n_\text{el}} x_e \,\tilde{\mathbf{K}}^{\mathrm{el}}(e),
    \label{eq:K_global}
\end{equation}
where $\tilde{\mathbf{K}}^{\mathrm{el}}(e)$ denotes the element matrix embedded at
the appropriate global rows and columns.  Fig.\,\ref{fig:stiffness_matrix_structure} visualizes this global matrix assembly. Supports are enforced by removing the corresponding rows and columns from $\mathbf{K}$, thereby imposing zero displacement on the respective DoF.

The discretized compliance corresponding to Eq.\,\eqref{eq:compliance_continuous} is given by
\begin{align}
    c(\mathbf{x}) &= \mathbf{u}^\top\mathbf{f} = \mathbf{u}^\top\mathbf{K}(\mathbf{x})\mathbf{u} \label{eq:compliance_literature}\\
    &=\mathbf{f}^\top\mathbf{K}^{-1}(\mathbf{x})\mathbf{f} \quad.
    \label{eq:compliance}
\end{align}
While the rightmost form in Eq.\,\eqref{eq:compliance_literature} is often used in literature \cite{sigmund_99_2001, andreassen_efficient_2011}, we use the equivalent form in Eq.\,\eqref{eq:compliance}, as it is easier to implement later on. Again without loss of generality, this assumes symmetric matrix $\mathbf{K}$.
The material volume fraction becomes
\begin{equation}
    V(\mathbf{x}) = \frac{1}{n_\text{el}}\sum_{e=1}^{n_\text{el}} x_e.
    \label{eq:volume_fraction}
\end{equation}
The discrete optimization problem is therefore
\begin{equation}
    \begin{split}
        \min_\mathbf{x}\ & c(\mathbf{x}), \\
        \text{subject to }& V(\mathbf{x}) = V_0.
    \end{split}
    \label{eq:FEM_topopt}
\end{equation}

\begin{figure}[h!] 
    \centering 
    \includegraphics[width=\columnwidth]{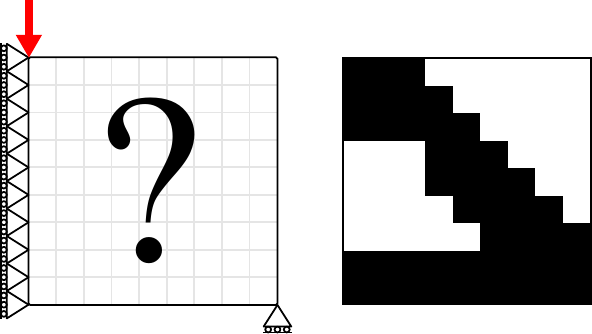} 
    \caption{Example solution for an MBB beam problem. The imposed boundary conditions and the square domain are displayed on the left. The horizontal DoFs of all nodes on the left boundary are fixed as well as the vertical DoF at the bottom-right corner. A single force pointing downwards is applied at the top-left corner. The right image shows a rounded optimized structure for $9\times9$ elements and a volume fraction of $V_0=1/2$ using the SIMP method outlined in Ref.\,\cite{andreassen_efficient_2011}.} 
    \label{fig:MBB_example} 
\end{figure}

Fig. \ref{fig:MBB_example} shows a solution of the MBB problem obtained with the SIMP method, which resembles the expected truss structure. Here, the volume fraction was set to $V_0=1/2$, meaning half of the design domain is filled with material.

\section{Quantum Algorithm}\label{sec:quantalg}
In the following, we introduce our fault-tolerant quantum algorithm for TO of the MBB beam benchmark problem.

\subsection{Quantum Topology Optimization Problem}
To solve the MBB problem on a quantum computer, we adapt the density-based TO problem in Eq.\,\eqref{eq:FEM_topopt}. We restrict the design variables to binary values,
\begin{equation}
    x_e \in \{0,1\}, \qquad x_e=1 \ \text{solid}, \quad x_e=0 \ \text{void}.
\end{equation}
This enables a natural representation of candidate solutions as computational basis states $\ket{x_1x_2\dots x_{n_\text{el}}}$ and is known as TO of binary structures \cite{picelli_101-line_2021}. In addition, we reformulate the problem as a satisfiability problem:
\begin{equation}
    \text{Find } \mathbf{x} \in \{0, 1\}^{n_\text{el}} \quad \text{such that} \quad
    \begin{cases}
    \begin{split}
        c(\mathbf{x}) &< c_{0}, \\
        V(\mathbf{x}) &= V_{0}.
    \end{split}
    \end{cases}
    \label{eq:satisfiability_problem}
\end{equation}
This makes it native to our quantum approach. By iteratively lowering the threshold value $c_0$, we can find the material distribution that minimizes compliance \cite{durr_quantum_1996}. However, we want to mention that comparing multiple candidate solutions allows to consider additional latent constraints, for example from the perspective of manufacturing.

\subsection{Grover's Algorithm}\label{sec:groverssearch}
Grover's algorithm is a well-known quantum algorithm for unstructured search \cite{grover_fast_1996, boyer_tight_1998}. 
With high probability, it returns one of $M$ marked items in an unstructured search space of size $N$ using $\mathcal{O}(\sqrt{N/M})$ oracle calls. Since the classical counterpart would require $\mathcal{O}({N/M})$ evaluations, Grover's algorithm gives a quadratic speedup. We use the search algorithm to solve the satisfiability problem in Eq.\,\eqref{eq:satisfiability_problem} and thus to find stiff structure designs $\mathbf{x}$.

\begin{figure}[h]
    \centering
    \begin{quantikz}[row sep=0.1cm, column sep=0.12cm]
        \lstick{$\ket{0}_\texttt{c}$} & \qwbundle{} & \gate{U_\text{init}} & & \gate{O} \gategroup[1, steps=2,style={dashed,rounded corners,fill=cyan!20, inner xsep=-1pt, inner ysep=-1pt},background]{$r$ repetitions}& \gate{D} &  & \meter{\mathbf{x}}
    \end{quantikz}
    \caption{Quantum circuit for Grover's algorithm. $U_\text{init}$ is used to prepare the superposition of the search space. The following $r$ iterations apply Oracle $O$ and Diffusion $D$ repeatedly, before finally the mesurement is conducted.}
    \label{fig:groverssearch}
\end{figure}
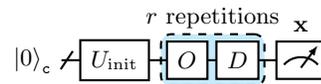

The quantum circuit is shown in Fig.\,\ref{fig:groverssearch}. 
We have a register $\texttt{c}$ containing $n_\texttt{c}=n_\text{el}$ qubits to represent structure configurations. The initialization $U_\text{init}$ prepares a superposition over the search space. If we search in the full space of size $N=2^{n_\text{el}}$, this operation simply becomes $U_\text{init} = H^{\otimes n_\text{el}}$. However, to enforce the volume constraint $V(\mathbf{x})=V_0$, we instead prepare the Dicke state \cite{gasieniec_deterministic_2019}
\begin{equation}
    U_\text{init}\ket{0}_\texttt{c} = \ket{D^{(n_\text{el})}_{k}}_\texttt{c} = \frac{1}{\sqrt{N}} \sum_{|\mathbf{x}|=k} \ket{\mathbf{x}}_\texttt{c},
\end{equation}
which is a superposition over the $N=\binom{n_\text{el}}{k}$ states with Hamming weight $k=\lfloor n_\text{el}V_0\rfloor$.

After the initialization, we have $r$ repetitions of Grover's oracle $O$ and Grover's diffusion operator $D$. $O$ marks every solution state as
\begin{equation}
    O\ket{\mathbf{x}} =
    \begin{cases}
        -\ket{\mathbf{x}}  &\text{for a solution }\mathbf{x},\\
        \ket{\mathbf{x}}  &\text{for any other  }\mathbf{x}.
    \end{cases}
    \label{eq:oracle}
\end{equation}
Grover's diffusion operator is given by
\begin{equation}
    D = -U_\text{init}S_0U_\text{init}^\dagger \quad\text{with}\quad S_0=\mathds{1}-2\ketbra{0}{0}.
    \label{eq:D}
\end{equation}  

\begin{figure}[h]
    \centering
    \includegraphics[width=0.65\columnwidth]{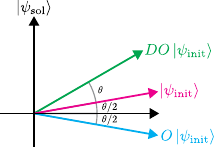}
    \caption{Geometrical illustration of a single Grover iteration.}
    \label{fig:grover_geometry}
\end{figure}

Fig.\,\ref{fig:grover_geometry} gives a geometric understanding of Grover's algorithm. 
We begin in the initial state ${U_\text{init}\ket{0} = \ket{\psi_\text{init}}}$. Grover's oracle reflects that state about the state orthogonal to our solution state $\ket{\psi_\text{sol}}$. The diffusion operator is another reflection about $\ket{\psi_\text{init}}$. By repeating these two operators, the state is rotated by $\theta$ toward the solution state. The angle $\theta$ is given by the expression
\begin{equation}
    \sin\left(\frac{\theta}{2}\right) = \braket{\psi_\text{sol}}{\psi_\text{init}} = \sqrt{\frac{M}{N}},
\end{equation}
and the number of repetitions $r$ should be chosen to maximize the overlap
\begin{equation}
    \bra{\psi_\text{sol}}(DO)^r\ket{\psi_\text{init}} = \sin\left(\left(r+\frac{1}{2}\right)\theta\right).
\end{equation}
Consequently, 
\begin{equation}
    r = \left\lfloor\frac{\pi}{4\arcsin\sqrt{M/N}}-\frac{1}{2}\right\rfloor \stackrel{M\ll N}{\approx} \left\lfloor\frac{\pi}{4}\sqrt{\frac{N}{M}}-\frac{1}{2}\right\rfloor.
    \label{eq:r}
\end{equation}
\subsection{Grover's Oracle}\label{sec:oracle}
Previously, Grover's oracle $O$ was introduced as marking solution states as shown in Eq.\,\eqref{eq:oracle}. In order to solve the satisfiability problem in Eq.\,\eqref{eq:satisfiability_problem}, the oracle should mark states where $c(\mathbf{x})<c_0$. When the configuration register is initialized as a Dicke state, the volume constraint $V(\mathbf{x})=V_0$ is already enforced and need not be checked by the oracle.

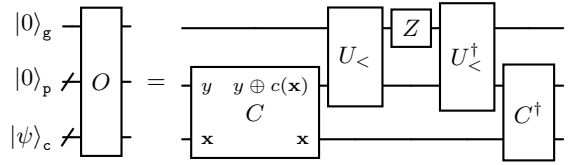
\begin{figure}[h]
    \centering
    \begin{quantikz}[row sep=0.1cm, column sep=0.12cm]
        \lstick{$\ket{0}_\texttt{g}$} & & \gate[3]{O} & \\[0.14cm]
        \lstick{$\ket{0}_\texttt{p}$} & \qwbundle{} & \ghost[2]{C} & \\[0.14cm]
        \lstick{$\ket{\psi}_\texttt{c}$} & \qwbundle{} & \ghost{C}& 
    \end{quantikz}
    =
    \begin{quantikz}[row sep=0.1cm, column sep=0.12cm]
        & & \gate[2]{U_{<}} & \gate{Z} & \gate[2]{U_{<}^\dagger} & & \\
        & \gate[2][1.7cm]{C} \gateinput{$y$}\gateoutput{$y \oplus c(\mathbf{x})$} & & & & \gate[2]{C^\dagger} & \\
        &\gateinput{$\mathbf{x}$}\gateoutput{$\mathbf{x}$} & & & & & 
    \end{quantikz}
    \caption{Quantum circuit for Grover's Oracle operation $O$.}
    \label{fig:groversoracle}
\end{figure}

We implement $O$ as depicted in Fig.\,\ref{fig:groversoracle}.
Here, the subroutine $C$ writes the compliance into an ancilla register $\texttt{p}$ as
\begin{equation}
    C\ket{0}_\texttt{p}\ket{\mathbf{x}}_\texttt{c} = \ket{c(\mathbf{x})}_\texttt{p}\ket{\mathbf{x}}_\texttt{c}.
    \label{eq:Cop}
\end{equation}
Thereafter, the operation $U_<$ flips the ancilla qubit $\texttt{g}$ if $c(\mathbf{x})<c_0$. This is followed by a $Z$ gate on register $\texttt{g}$ producing the phase flip of Eq.\,\eqref{eq:oracle}, and we uncompute previous operations.

\subsection{Compliance Computation}\label{sec:compliancecomputation}
Next, we dive into our implementation of $C$ computing the compliance as defined in Eq.\,\eqref{eq:Cop}. By recalling Eq.\,\eqref{eq:compliance}, we can express $c(\mathbf{x})$ as
\begin{equation}
    \tfrac{1}{\alpha}\,c(\mathbf{x}) = \bra{\mathbf{x}}_\texttt{c}\bra{\mathbf{f}}_\texttt{d}\bra{0}_\texttt{qlvzb}U_{\mathbf{K}^{-1}}\ket{0}_\texttt{qlvzb}\ket{\mathbf{f}}_\texttt{d}\ket{\mathbf{x}}_\texttt{c}
    \label{eq:expectationvalue}
\end{equation}
up to a known normalization constant $\alpha$ as described later.
Here, $U_{\mathbf{K}^{-1}}$ is a unitary operator providing the inverse global stiffness matrix $\mathbf{K}^{-1}$ as a block-encoding for the configuration $\ket{\mathbf{x}}_\texttt{c}$. The registers \texttt{qlvzb} serve as ancillas for the block-encoding, and the Hilbert space of the data register \texttt{d} carries $\mathbf{f}$ and $\mathbf{K}$. Further details on the block-encoding follow in the next two sections. 

\begin{figure}[h!]
    \centering
    \begin{quantikz}[row sep=0.1cm, column sep=0.12cm]
        \lstick{$\ket{0}_\texttt{h}$} &                      & \gate[4]{A} & \\[0.02cm]
        \lstick{$\ket{0}_\texttt{qlvzb}$} & \qwbundle{}      & \ghost{H} &\\[-0.18cm]
        \lstick{$\ket{0}_\texttt{d}$} & \qwbundle{}          & \ghost{V_{\mathbf{f}}} & \\[-0.18cm]
        \lstick{$\ket{\psi}_\texttt{c}$} & \qwbundle{} & \ghost{H} & \\
    \end{quantikz}
    =
    \begin{quantikz}[row sep=0.1cm, column sep=0.12cm]
        & \gate{H} & \ctrl{1} & \gate{H} & \\
        & \ghost{H} & \gate[3]{U_{\mathbf{K}^{-1}}} & &\\[-0.2cm]
        & \gate{V_\mathbf{f}} & & & \\[-0.2cm]
        & \ghost{H} & & & \\
    \end{quantikz}
    \caption{Quantum circuit diagram of the Hadamard Test $A$.}
    \label{fig:hadamardtest}
\end{figure}
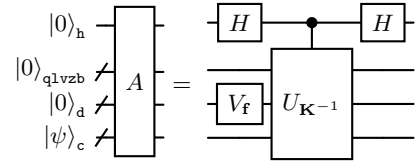

\begin{figure}[h!]
    \centering
    \begin{quantikz}[row sep=0.1cm, column sep=0.12cm]
        \lstick{$\ket{0}_\texttt{p}$} &                      & \gate[8]{C} & \\[-0.08cm]
        \lstick{\vdots } &  \setwiretype{n}                  & \ghost{H} & \\
        \lstick{$\ket{0}_\texttt{p}$} &                      & \ghost{H} & \\[0.04cm]
        \lstick{$\ket{0}_\texttt{p}$} &                      & \ghost{H} & \\[0.1cm]
        \lstick{$\ket{0}_\texttt{h}$} &                      & \ghost{H} &\\[-0.12cm]
        \lstick{$\ket{0}_\texttt{qlvzb}$} & \qwbundle{}      & \ghost{H} & \\[-0.2cm]
        \lstick{$\ket{0}_\texttt{d}$} & \qwbundle{}          & \ghost{H} & \\[-0.12cm]
        \lstick{$\ket{\psi}_\texttt{c}$} & \qwbundle{} & \ghost{H} & \\
    \end{quantikz}
    =
    \begin{quantikz}[row sep=0.1cm, column sep=0.12cm]
        & \gate{H} & & & & \ctrl{4} &  \gate[4]{\text{QFT}^\dagger} & \\[-0.18cm]
        \setwiretype{n} & \vdots & &  & \iddots & & & \\
        & \gate{H} & & \ctrl{2} & & & & \\
        & \gate{H} & \ctrl{1} & & & & & \\
        & \gate[4]{A} & \gate[4]{G} & \gate[4]{G^2} & \ \ldots \ & \gate[4]{G^{2^{n_\texttt{p}-1}}} & &\\[-0.2cm]
        & \ghost{H} & & & &\ \ldots \ & & \\[-0.2cm]
        & \ghost{H} & & & &\ \ldots \ & & \\[-0.2cm]
        & \ghost{H} & & & &\ \ldots \ & & \\
    \end{quantikz}
    \caption{Quantum circuit for Quantum Amplitude Estimation within the compliance computation operation $C$. The $\text{QFT}^\dagger$ block represents an inverse Quantum Fourier Transform (QFT).}
    \label{fig:QAE}
\end{figure}
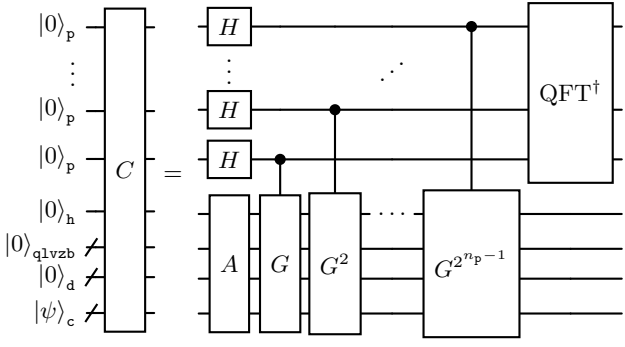

\begin{figure}[h!]
    \centering
    \begin{quantikz}[row sep=0.1cm, column sep=0.12cm]
        \lstick{$\ket{0}_\texttt{h}$} &                      & \gate[4]{G} &\\[-0.18cm]
        \lstick{$\ket{0}_\texttt{qlvzb}$} & \qwbundle{}      & \ghost{H} &\\[-0.22cm]
        \lstick{$\ket{0}_\texttt{d}$} & \qwbundle{}          & \ghost{H} &\\[-0.18cm]
        \lstick{$\ket{\psi}_\texttt{c}$} & \qwbundle{} & \ghost{H} & \\
    \end{quantikz}
    =
    \begin{quantikz}[row sep=0.1cm, column sep=0.12cm]
        & \gate{S_0} & \gate[4]{A^\dagger} & \gate[4]{-S_0} & \gate[4]{A} &\\[-0.2cm]
        & \ghost{H} & & & &\\[-0.2cm]
        & \ghost{H} & & & &\\[-0.2cm]
        & \ghost{H} & & & & \\
    \end{quantikz}
    \caption{Grover Operator $G$ within the Quantum Amplitude Estimation circuit.}
    \label{fig:groveroperator}
\end{figure}
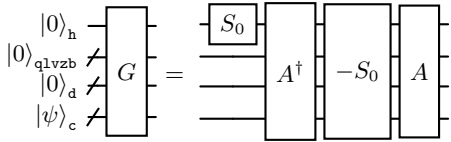

To compute the expectation value in Eq.\,\eqref{eq:expectationvalue}, we use a Hadamard test \cite{lin_lecture_2022} $A$ with an ancilla qubit $\texttt{h}$ as shown in Fig.\,\ref{fig:hadamardtest}.
The state preparation $V_\mathbf{f}\ket{0}_\texttt{d} = \ket{\mathbf{f}}$ loads the normalized force vector. The action of $A$ is
\begin{equation}
\begin{split}
    A\ket{0}_\texttt{h}&\ket{0}_\texttt{qlvzb}\ket{0}_\texttt{d}\ket{\mathbf{x}}_\texttt{c} \\
    =&\frac{1}{2}\Bigl[\ket{0}_\texttt{h}\left(\mathds{1} + U_{\mathbf{K}^{-1}}\right)\ket{0}_\texttt{qlvzb}\ket{\mathbf{f}}_\texttt{d}\ket{\mathbf{x}}_\texttt{c}\\
    &+ \ket{1}_\texttt{h}\left(\mathds{1} - U_{\mathbf{K}^{-1}}\right)\ket{0}_\texttt{qlvzb}\ket{\mathbf{f}}_\texttt{d}\ket{\mathbf{x}}_\texttt{c}\Bigr].
\end{split}
\end{equation}
We embed $A$ into a QAE \cite{lomonaco_quantum_2002} to estimate the absolute amplitude $\abs{h_0}$ of $\ket{0}_\texttt{h}$. The corresponding quantum circuit is given in Fig.\,\ref{fig:QAE}.
In general, QAE is a quantum phase estimation circuit \cite{nielsen_quantum_2012} with the Grover operator $G$ as central unitary. The definition of $G$ is shown in Fig.\,\ref{fig:groveroperator}. Note that $G$ is unrelated to Grover's algorithm discussed in Section\,\ref{sec:groverssearch}.
Consequently, the QAE routine estimates the absolute amplitude
\begin{equation}
\resizebox{\columnwidth}{!}{$
\begin{aligned}
    \lvert h_0\rvert
    &= \sqrt{\tfrac{1}{2}+\tfrac{1}{2}\Re{
        \bra{\mathbf{x}}_{\texttt{c}}\bra{\mathbf{f}}_{\texttt{d}}\bra{0}_{\texttt{qlvzb}}\,
        U_{\mathbf{K}^{-1}}\,
        \ket{0}_{\texttt{qlvzb}}\ket{\mathbf{f}}_{\texttt{d}}\ket{\mathbf{x}}_{\texttt{c}}
    }}\\
    &= \sqrt{\tfrac{1}{2}+\tfrac{1}{2\alpha}\,c(\mathbf{x})}
\end{aligned}
$}
\label{eq:h0}
\end{equation}
and writes the phase
\begin{equation}
    \theta(\mathbf{x}) = \pm\tfrac{1}{\pi}\arcsin\left(\sqrt{\tfrac{1}{2}+\tfrac{1}{2\alpha}c(\mathbf{x})}\right)
    \label{eq:phase_cost}
\end{equation}
into the phase register $\texttt{p}$. Thus, our implementation of $C$ differs from the ideal definition in Eq.\,\eqref{eq:Cop} and computes
\begin{equation}
    C\ket{0}_\texttt{p}\ket{\mathbf{x}}_\texttt{c} = \frac{1}{\sqrt{2}}\left(\ket{\theta(\mathbf{x})}_\texttt{p} + \ket{-\theta(\mathbf{x})}_\texttt{p}\right)\ket{\mathbf{x}}_\texttt{c}.
\end{equation}
We consider this as a technical detail, since the $\pm$ branches of $\theta(c(\mathbf{x})) = \theta(\mathbf{x)}$ are strictly increasing bijections of $c(\mathbf{x})$, and we equivalently compare ${\abs{\theta(\mathbf{x})}<\theta_0=\arcsin\left(\sqrt{\tfrac{1}{2}+\tfrac{1}{2\alpha}c_0}\right)/\pi}$ inside $U_<$.

\subsection{Matrix Inversion}\label{sec:qsvt}
In this section, we explain how to implement the unitary $U_{\mathbf{K}^{-1}}$ block-encoding the inverse global stiffness matrix $\mathbf{K}^{-1}(\mathbf{x})$ for a configuration $\mathbf{x}$. We employ the QSVT algorithm \cite{martyn_grand_2021, gilyen_quantum_2019} to invert the global stiffness matrix $\mathbf{K}(\mathbf{x})$. Therefore, we assume access to the corresponding block-encoding $U_{\mathbf{K}}$ which is detailed in the next section. QSVT applies a polynomial transformation to the singular values of the block-encoded matrix without explicitly accessing the singular value decomposition. In our case, this matrix is $\mathbf{K}$ up to a scaling constant $\beta$. Hence, 
\begin{equation}
    \frac{1}{\beta}\mathbf{K} = \mathbf{V}\mathbf{\Sigma}\mathbf{W}^\dagger \overset{\text{QSVT}}{\mapsto} \mathbf{V}P(\mathbf{\Sigma})\mathbf{W}^\dagger,
\end{equation}
where $P(\cdot)$ is the QSVT polynomial, $\mathbf{\Sigma}$ contains the singular values $\sigma_i\in[0,1]$ in its diagonal, and $\mathbf{V}, \mathbf{W}$ are the unitaries of the singular value decomposition (SVD). Since $\mathbf{K}=\mathbf{K^\top}$, the SVD is equivalent to the eigenvalue decomposition and $\mathbf{V}=\mathbf{W}$, however we are not restricted to symmetric matrices in general. For matrix inversion, the QSVT polynomial approximates the reciprocal function ${P_\text{inv}(x)\approx \gamma/x}$ on the interval $x\in[\mu,1]$, where $\mathcal{O}(\gamma)=\mathcal{O}(\mu)$ ensures that the function values lie within $[-1, 1]$, and $\mu$ is the smallest relevant singular value. Consequently,
\begin{equation}
    \frac{1}{\beta}\mathbf{K} = \mathbf{V}\mathbf{\Sigma}\mathbf{V}^\dagger \overset{\text{QSVT}}{\mapsto} \mathbf{V}P_\text{inv}(\mathbf{\Sigma})\mathbf{V}^\dagger \approx \gamma\beta\mathbf{K}^{-1}.
\end{equation}

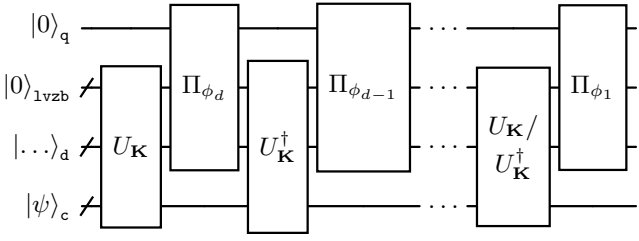
\begin{figure}[h!]
    \centering
    \begin{adjustbox}{width=\columnwidth}
    \begin{quantikz}[row sep=0.1cm, column sep=0.12cm]
        \lstick{$\ket{0}_\texttt{q}$} & & &\gate[3]{\Pi_{\phi_{d}}}& & \gate[3]{\Pi_{\phi_{d-1}}} & \ \ldots \ & & \gate[3]{\Pi_{\phi_1}} & \\
        \lstick{$\ket{0}_\texttt{lvzb}$} & \qwbundle{} & \gate[3]{U_\mathbf{K}} & & \gate[3]{U_\mathbf{K}^\dagger} & & \ \ldots \ & \gate[3, disable auto height]{\shortstack{$U_\mathbf{K}$\text{/}\\ $U_\mathbf{K}^\dagger$}} & & \\
        \lstick{$\ket{\dots}_\texttt{d}$} & \qwbundle{} & & &  & & \ \ldots \ & & & \\
        \lstick{$\ket{\psi}_\texttt{c}$} & \qwbundle{} & & & & & \ \ldots \ & & & 
    \end{quantikz}
    \end{adjustbox}
    \caption{Quantum circuit realizing a block-encoding of the inverse stiffness matrix $\mathbf{K}^{-1}$ via Quantum Singular Value Transform.}
    \label{fig:qsvt}
\end{figure}

The QSVT algorithm is defined as an alternating sequence of the block-encoding $U_\mathbf{K}$ (or its conjugate transpose) and projector-controlled phase-shifts ${\Pi_{\phi_k}=e^{i\phi_k(2\Pi-\mathds{1})}}$. Here, $\Pi$ is the projector to the block within $U_\mathbf{K}$ corresponding to $\mathbf{K}$, and $\{\phi_k\}$ are phase angles. The QSVT theorem states that it is possible to find $d$ angles $\{\phi_1, \phi_2, \dots, \phi_d\}$ that implement any polynomial $P(\cdot)$ with $\deg(P)\leq d$, parity $d\,\text{mod}\,2$, and specific norm to preserve normalization. We discuss the choice of the polynomial approximation and the angle finding algorithms in Section\,\ref{sec:qsvt_poly}. Consequently, we have
\begin{align}
\begin{split}
    U_{\mathbf{K}^{-1}} &= 
    \begin{cases}
        \Pi_{\phi_1}U_\mathbf{K}\prod\limits_{k=1}^{(d-1)/2}\Pi_{\phi_{2k}}U_\mathbf{K}^\dagger\Pi_{\phi_{2k+1}}U_\mathbf{K}\text{,}&\text{for odd }d \\
        \prod\limits_{k=1}^{d/2}\Pi_{\phi_{2k-1}}U_\mathbf{K}^\dagger\Pi_{\phi_{2k}}U_\mathbf{K}\text{,}&\text{for even }d
    \end{cases}\\
    &=
    \begin{pmatrix}
        \begin{array}{
          w{c}{\widthof{$\mathbf{V}P_\text{inv}(\mathbf{\Sigma})\mathbf{V}^\dagger$}}
          w{c}{\widthof{$\mathbf{V}P_\text{inv}(\mathbf{\Sigma})\mathbf{V}^\dagger$}}
          w{c}{\widthof{$\mathbf{V}P_\text{inv}(\mathbf{\Sigma})\mathbf{V}^\dagger$}}
          w{c}{\widthof{$\mathbf{V}P_\text{inv}(\mathbf{\Sigma})\mathbf{V}^\dagger$}}
        }
        \mathbf{V}P_\text{inv}(\mathbf{\Sigma})\mathbf{V}^\dagger & *\\
        * & *
    \end{array}
    \end{pmatrix}.
\end{split}
\label{eq:U_K_inverse}
\end{align}
The corresponding quantum circuit looks like Fig.\,\ref{fig:qsvt}, where we use the ancilla qubit $\texttt{q}$ to implement $\Pi_{\phi_k}$, and $U_\mathbf{K}$ depends on the block-encoding ancillas $\texttt{lvzb}$, the data register $\texttt{d}$, and the configuration register $\texttt{c}$.

\begin{figure}[h!]
    \centering
    \begin{quantikz}[row sep=0.2cm, column sep=0.12cm]
        \lstick{$\ket{0}_\texttt{q}$} & & \targ{} \gategroup[7,steps=4,style={dashed,rounded corners,fill=cyan!20, inner xsep=-1pt, inner ysep=-1pt},background]{fix DoFs} & \targ{}    & \targ{}    & \targ{}    & \targ{}    & \gate{e^{-i\phi_d Z}} & \targ{} & \targ{} \gategroup[7,steps=4,style={dashed,rounded corners,fill=cyan!20, inner xsep=-1pt, inner ysep=-1pt},background]{fix DoFs} & \targ{} & \targ{} & \targ{} &\\[-0.3cm]
        \lstick{$\ket{0}_\texttt{lvzb}$} & \qwbundle{} & \octrl{-1} & \octrl{-1} & \octrl{-1} & \octrl{-1} & \octrl{-1} & \ghost{H} & \octrl{-1} & \octrl{-1} & \octrl{-1} & \octrl{-1} & \octrl{-1} &\\[-0.3cm]
        \lstick[5]{$\ket{\dots}_\texttt{d}$} & & \octrl{-1} & \octrl{-1} & \octrl{-1} & \ctrl{-1} & & \ghost{H} & & \ctrl{-1} & \octrl{-1} & \octrl{-1} & \octrl{-1} &\\[-0.3cm]
         & & \octrl{-1} & \octrl{-1} & \octrl{-1} & \octrl{-1} & & \ghost{H} & & \octrl{-1} & \octrl{-1} & \octrl{-1} & \octrl{-1} &\\[-0.3cm]
         & & \octrl{-1} & \octrl{-1} & \ctrl{-1} & \octrl{-1} & & \ghost{H} & & \octrl{-1} & \ctrl{-1} & \octrl{-1} & \octrl{-1} &\\[-0.3cm]
         & & \octrl{-1} & \ctrl{-1} & \octrl{-1} & \octrl{-1} & & \ghost{H} & & \octrl{-1} & \octrl{-1} & \ctrl{-1} & \octrl{-1} &\\[-0.3cm]
         & & \octrl{-1} & \octrl{-1} & \octrl{-1} & \ctrl{-1} & & \ghost{H} & & \ctrl{-1} & \octrl{-1} & \octrl{-1} & \octrl{-1} &
    \end{quantikz}
    \caption{Exemplary circuit implementation of the projector-controlled phase-shift $\Pi_{\phi_d}$. The blue shaded operations remove the fixed DoFs with indices 1, 3, 4, and 18 (least significant bit on the bottom).}
    \label{fig:pcphase}
\end{figure}
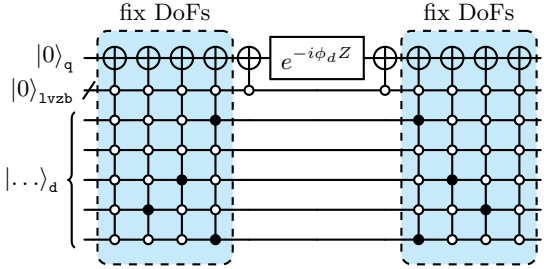

To implement $\Pi_{\phi_k}$, one typically uses the projector $\Pi = \ketbra{0}{0}_{\texttt{lvzb}}\otimes\mathds{1}_\texttt{d}$ pointing to the top-left block of the unitary. We modify the projector to remove the rows and columns corresponding to the fixed DoFs 
\begin{equation}
    \Pi = \ketbra{0}{0}_{\texttt{lvzb}}\otimes \left(\mathds{1}_\texttt{d}- \sum_{\text{\shortstack{fixed DoFs} }i} \ketbra{i}{i}\right).
\end{equation}
Fig.\,\ref{fig:pcphase} shows the circuit implementation of $\Pi_{\phi_k}$ (up to a global phase) for the example $2\times2$ elements example, where we remove the DoFs with indices 1, 3, 4, and 18 to encode horizontal support on the left boundary and vertical support at the bottom-right corner corresponding to the example depicted in fig.\,\ref{fig:MBB_example}.

\subsection{Matrix Block-Encoding}\label{sec:mbe}
\begin{figure}[h!]
    \centering
    \includegraphics[width=\columnwidth]{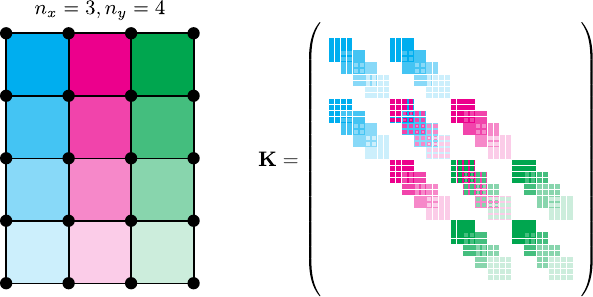}
    \caption{Element-wise contributions to the global stiffness matrix
             for a $3\times4$ domain. The element's colors match their contribution.}
    \label{fig:stiffness_matrix_structure}
\end{figure}

The QSVT circuit for matrix inversion is based on $U_\mathbf{K}$ block-encoding the global stiffness matrix $\mathbf{K}$ in the top-left block
\begin{equation}
    U_{\mathbf{K}} = \frac{1}{\beta}\begin{pmatrix}
        \begin{array}{
          w{c}{\widthof{$\mathbf{K}$}}
          w{c}{\widthof{$\mathbf{K}$}}
          w{c}{\widthof{$\mathbf{K}$}}
          w{c}{\widthof{$\mathbf{K}$}}
        }
        \mathbf{K} & *\\
        * & *
    \end{array}
    \end{pmatrix},
    \label{eq:U_K}
\end{equation}
where $\beta$ is a scaling constant ensuring $\norm{\mathbf{K}/\beta} \leq 1$. Furthermore, $U_\mathbf{K}$ depends on the structure design $\mathbf{x}$ stored in the configuration register $\ket{\mathbf{x}}_\texttt{c}$. From Eq.\,\eqref{eq:K_global}, we know that $\mathbf{K}$ is the sum of the element stiffness matrices $\mathbf{K}^\text{el}$ embedded into the global DoF space. Fig.\,\ref{fig:stiffness_matrix_structure} illustrates this assembly for a $n_x=3$, $n_y=4$ domain. 
The matrix $\mathbf{K}$ is sparse, banded, and displays a repeating block pattern due to the uniform element topology and global DoF ordering.
Each element $e$ contributes a transformed element stiffness matrix $\tilde{\mathbf{K}}^\text{el}(e)$ to $\mathbf{K}$. $\tilde{\mathbf{K}}^\text{el}(e)$ is $\mathbf{K}^\text{el}$ with an offset $\Delta(e)$ in both rows and columns and a gap of size $2(n_y-1)$ splitting the $8\times8$ matrix into four $4\times4$ blocks. This transformation is visualized by 
\begin{equation}
    \mathbf{K}^\text{el}=\adjincludegraphics[valign=c]{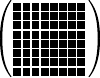} \mapsto \tilde{\mathbf{K}}^\text{el}(e) = \adjincludegraphics[valign=c]{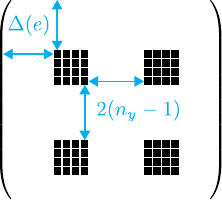},
    \label{eq:Ketilde}
\end{equation}
where $\tilde{\mathbf{K}}^\text{el}(e)$ has size $n_\text{DoF}\times n_\text{DoF}$. This transformation embeds each local matrix into the global domain, aligning the DoFs with the column-wise global ordering. The diagonal offset is
\begin{equation}
    \Delta(x,y) = 2\left[(x-1)(n_y+1)+(y-1)\right]
\end{equation}
with the element coordinates $1\leq x \leq n_x$ and $1\leq y \leq n_y$. If we count the elements column-wise, we have
\begin{equation}
    e = (x-1)n_y + y \quad \text{with} \quad e=1,\dots, n_\text{el}
\end{equation}
and conversely
\begin{equation}
    x = \left\lfloor \frac{e-1}{n_y} \right\rfloor +1 \quad \text{and} \quad y = (e-1)\bmod n_y+1.
\end{equation}
Consequently, the offset for an element $e$ is
\begin{equation}
    \Delta(e) = 2\left(e-1+\left\lfloor \frac{e-1}{n_y} \right\rfloor\right).
\end{equation}

Based on the matrix structure, we implement the block-encoding of $\mathbf{K}$ using the following six-step strategy.
\subsubsection{Block-encoding of the element stiffness matrix}
We begin by block-encoding the element stiffness matrix $\mathbf{K}^\text{el}/\delta$, where $\delta$ is a scaling constant. Since $\mathbf{K}^\text{el}$ has constant size, this block-encoding can be implemented in $\mathcal{O}(1)$ steps using for example the Quantum Shannon Decomposition \cite{shende_synthesis_2006} to decompose 
\begin{equation}
U_{\mathbf{K}^\text{el}} = \frac{1}{\delta}
\begin{pmatrix}
    \mathbf{K}^\text{el} & \sqrt{\delta^2\mathds{1}-\mathbf{K}^{\text{el}^2}} \\
    \sqrt{\delta^2\mathds{1}-\mathbf{K}^{\text{el}^2}} & \mathbf{K}^\text{el} 
\end{pmatrix},
\label{eq:U_Kel}
\end{equation}
into basis gates with block-encoding ancilla qubit $\texttt{b}$.
\subsubsection{Zero-padding to global size}\label{sec:zero_padding}
To embed $\mathbf{K}^\text{el}$ into the global DoF space, we pad it with $n_\text{DoF}-8$ zeros. Therefore, we expand the data register $\texttt{d}$ by 
\begin{equation}
    d_\text{z} = \left\lceil\log_2\frac{n_\text{DoF}}{8}\right\rceil
\end{equation}
qubits. To remove non-zero entries originating from implicit identity gates on these $d_\text{z}$ qubits, an operation $F$ flips an additional block-encoding ancilla qubit $\texttt{z}$ whenever the added $d_\text{z}$ qubits differ from $\ket{0\cdots0}$.

\subsubsection{Gap insertion}
\begin{figure}[hbt!]
    \begin{equation*}
    \begin{quantikz}[row sep=0.1cm, column sep=0.2cm]
        \lstick{$\ket{f_0}_\texttt{d}$} & \gate[2]{P_\text{gap}} &\\
        \lstick{$\ket{f_1\dots f_{n_\text{e}}}_\texttt{d}$} & & 
    \end{quantikz}
        =
    \begin{quantikz}[row sep=0.1cm, column sep=0.2cm]
         & \gate[2]{P_{-4}} & \octrl{1} & \gate[2]{P_{+4}} &\\
         & & \gate{P_{+2(n_y-1)}} & &
    \end{quantikz}
    \end{equation*}
    \caption{Quantum circuit for the permutation operation $P_\text{gap}$ introducing the gap of size $2(n_y-1)$ as shown in Eq.\,\eqref{eq:Ketilde}.}
    \label{fig:insert_gap}
\end{figure}
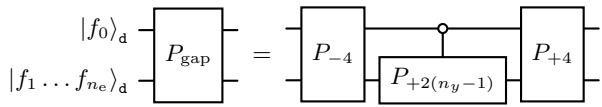

Due to the column-wise ordering of global DoFs, a gap of size $2(n_y-1)$ must separate the left and right nodes of each element. We implement this by permuting matrix indices with a permutation unitary $P_\text{gap}$ acting on the data register,
\begin{equation}
P_\text{gap}=P_{+4}\left(P_{+2(n_y-1)}\;\text{(controlled on MSB)}\right)P_{-4},
\end{equation}
where $P_{\pm k}$ adds/subtracts $k$ to/from the binary index (modulo the register range) and are realized with standard reversible adders. Here, $P_{-4}$ shifts the beginning of the gap to the zero-index. The gap is inserted via $P_{+2(n_y-1)}$, which is controlled on the most significant bit and acts on the remaining bits, thereby shifting only the first half of the matrix entries. $P_{+4}$ is reversing the initial shift. Fig.\,\ref{fig:insert_gap} shows the quantum circuit for $P_\text{gap}$.

\subsubsection{Shift by global offset $\Delta(e)$}
Each element's contribution is shifted by $\Delta(e)$ into its global position using permutations $P_{\pm\Delta(e)}$, yielding
\begin{equation}
    U_{\tilde{\mathbf{K}}^\text{el}(e)} = P_{+\Delta(e)}P_\text{gap}FU_{\mathbf{K}^\text{el}}P_\text{gap}^\dagger P_{-\Delta(e)}.
    \label{eq:U_tildeKel}
\end{equation}
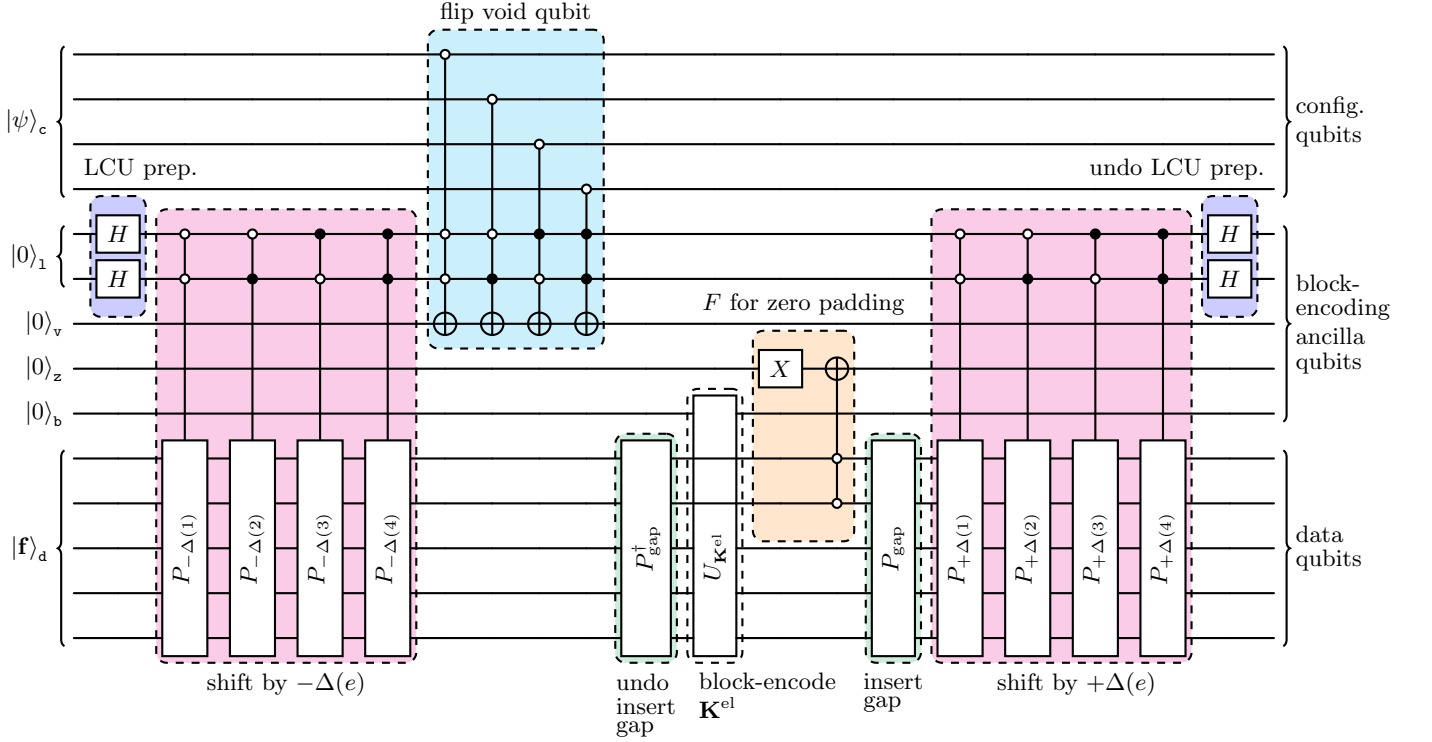
\begin{figure*}
    \centering
    \begin{quantikz}[row sep=0.1cm, column sep=0.3cm]
        \lstick[4]{$\ket{\psi}_\texttt{c}$} & \ghost{H} & & & & & \octrl{4}
        \gategroup[7,steps=4,style={dashed,rounded corners,fill=mycyan!20, inner xsep=-1pt, inner ysep=-1pt},background]{flip void qubit} & & & & & & & & & & & & & & \rstick[4]{\shortstack[l]{config.\\qubits}}\\
        & \ghost{H} & & & & & & \octrl{3} & & & & & & & & & & & & & \\
        & \ghost{H} & & & & & & & \octrl{2} & & & & & & & & & & & & \\
        & \ghost{H} & & & & & & & & \octrl{1} & & & & & & & & & & & \\
        \lstick[2]{$\ket{0}_\texttt{l}$} & \gate{H} 
        \gategroup[2,steps=1,style={dashed,rounded corners,fill=blue!20, inner xsep=-1pt, inner ysep=4pt},background,label style={label position=above,anchor=north,yshift=0.4cm,xshift=0.3cm}]{{LCU prep.}}
        & \octrl{1} 
        \gategroup[10,steps=4,style={dashed,rounded corners,fill=mymagenta!20, inner xsep=-1pt, inner ysep=-1pt},background,label style={label position=below,anchor=north,yshift=-0.2cm}]{shift by $-\Delta(e)$}
        & \octrl{1} & \ctrl{1} & \ctrl{1} & \octrl{1} & \octrl{1} & \ctrl{1} & \ctrl{1} & & & & & & \octrl{1} 
        \gategroup[10,steps=4,style={dashed,rounded corners,fill=mymagenta!20, inner xsep=-1pt, inner ysep=-1pt},background,label style={label position=below,anchor=north,yshift=-0.2cm,xshift=0.2cm}]{shift by $+\Delta(e)$}
        & \octrl{1} & \ctrl{1} & \ctrl{1} & \gate{H} 
        \gategroup[2,steps=1,style={dashed,rounded corners,fill=blue!20, inner xsep=-1pt, inner ysep=4pt},background,label style={label position=above,anchor=north,yshift=0.4cm,xshift=-0.7cm}]{undo LCU prep.}
        & \rstick[5]{\shortstack[l]{block-\\encoding\\ancilla\\qubits}}\\
         & \gate{H} & \octrl{4} & \ctrl{4} & \octrl{4} & \ctrl{4} & \octrl{1} & \ctrl{1} & \octrl{1} & \ctrl{1} & & & & & & \octrl{4} & \ctrl{4} & \octrl{4} & \ctrl{4} & \gate{H} & \\
        \lstick{$\ket{0}_\texttt{v}$} & \ghost{H} & & & & & \targ{} & \targ{} & \targ{} & \targ{} & & & & & & & & & & & \\
        \lstick{$\ket{0}_\texttt{z}$} & \ghost{H} & & & & & & & & & & & \gate{X} 
        \gategroup[4,steps=2,style={dashed,rounded corners,fill=orange!20, inner xsep=-1pt, inner ysep=4pt},background,label style={label position=above,anchor=north,yshift=0.4cm}]{$F$ for zero padding}
        & \targ{} & & & & & & & \\
        \lstick{$\ket{0}_\texttt{b}$} & \ghost{H} & & & & & & & & & & \gate[wires=6, label style={yshift=-0.4cm}, style={fill opacity=0}]{\text{\rotatebox[origin=c]{90}{\makebox[0.2cm][c]{$U_{\mathbf{K}^\text{el}}$}}}}
        \gategroup[6,steps=1,style={dashed,rounded corners,fill=white!0, inner xsep=-1pt, inner ysep=-1pt},background,label style={label position=below,anchor=north,yshift=-0.2cm, xshift=0.7cm}]{\shortstack[l]{block-encode\\$\mathbf{K}^\text{el}$}}
        & & & & & & & & & \\
        \lstick[5]{$\ket{\mathbf{f}}_\texttt{d}$}  & \ghost{H} & \gate[wires=5]{\text{\rotatebox[origin=c]{90}{\makebox[0.2cm][c]{$P_{-\Delta(1)}$}}}} & \gate[wires=5]{\text{\rotatebox[origin=c]{90}{\makebox[0.2cm][c]{$P_{-\Delta(2)}$}}}} & \gate[wires=5]{\text{\rotatebox[origin=c]{90}{\makebox[0.2cm][c]{$P_{-\Delta(3)}$}}}} & \gate[wires=5]{\text{\rotatebox[origin=c]{90}{\makebox[0.2cm][c]{$P_{-\Delta(4)}$}}}} & & & & & \gate[wires=5]{\text{\rotatebox[origin=c]{90}{\makebox[0.2cm][c]{$P_\text{gap}^\dagger$}}}} 
        \gategroup[5,steps=1,style={dashed,rounded corners,fill=mygreen!20, inner xsep=-1pt, inner ysep=-1pt},background,label style={label position=below,anchor=north,yshift=-0.209cm}]{\shortstack[l]{undo\\insert\\gap}}
        & \linethrough & & \octrl{-2} & \gate[wires=5]{\text{\rotatebox[origin=c]{90}{\makebox[0.2cm][c]{$P_\text{gap}$}}}} 
        \gategroup[5,steps=1,style={dashed,rounded corners,fill=mygreen!20, inner xsep=-1pt, inner ysep=-1pt},background,label style={label position=below,anchor=north,yshift=-0.209cm}]{\shortstack[l]{insert\\gap}}
        & \gate[wires=5]{\text{\rotatebox[origin=c]{90}{\makebox[0.2cm][c]{$P_{+\Delta(1)}$}}}} & \gate[wires=5]{\text{\rotatebox[origin=c]{90}{\makebox[0.2cm][c]{$P_{+\Delta(2)}$}}}} & \gate[wires=5]{\text{\rotatebox[origin=c]{90}{\makebox[0.2cm][c]{$P_{+\Delta(3)}$}}}} & \gate[wires=5]{\text{\rotatebox[origin=c]{90}{\makebox[0.2cm][c]{$P_{+\Delta(4)}$}}}} & & \rstick[5]{\shortstack[l]{data\\qubits}} \\
          & \ghost{H} & & & & & & & & & & \linethrough & & \octrl{-1} & & & & & & & \\
          & \ghost{H} & & & & & & & & & & & & & & & & & & & \\
          & \ghost{H} & & & & & & & & & & & & & & & & & & & \\
          & \ghost{H} & & & & & & & & & & & & & & & & & & & 
    \end{quantikz}
    \caption{Quantum circuit representation of the operator $U_\mathbf{K}$ block-encoding the global stiffness matrix $\mathbf{K}$ up to a scaling factor for a domain with four elements.}
    \label{fig:block_encoding}
\end{figure*}
\subsubsection{Sum of all $U_{\tilde{\mathbf{K}}^\text{el}(e)}$}
We sum over all element contributions using the Linear Combination of Unitaries (LCU) method \cite{childs_hamiltonian_2012, sunderhauf_block-encoding_2024}. Therefore, we prepare a block-encoding ancilla register $\texttt{l}$ of $\lceil\log_2n_\text{el}\rceil$ qubits in a uniform superposition using Hadamards. This is sufficient, because using the fully structured mesh, all elements contribute equally to the global stiffness matrix. This is followed by the unitaries $U_{\tilde{\mathbf{K}}^\text{el}(e)}$ controlled on the corresponding binary index $e-1$ in $\texttt{l}$. Finally, Hadamards on $\texttt{l}$ uncompute the initial superposition. This circuit block-encodes the sum
\begin{equation}
    \frac{1}{n_\text{el}}\sum_{e=1}^{n_\text{el}}U_{\tilde{\mathbf{K}}^\text{el}(e)}.
    \label{eq:lcusum}
\end{equation}
Because only the outer permutations in Eq.\,\eqref{eq:U_tildeKel} depend on $e$, the inner operations (gap insertion, zero-padding, and $U_{\mathbf{K}^\text{el}}$) can be shared across all elements.
\subsubsection{Conditioning on material distribution $\mathbf{x}$}
To reflect the actual material distribution $\mathbf{x}$, we introduce a void ancilla $\texttt{v}$ flipped when $x_e=0$ and when the $\texttt{l}$ register from step 5 is in state $\ket{e-1}_\texttt{l}$. Thereby, we set the contribution of an empty element to zero and $\mathbf{K}$ is constructed only from filled elements and depends coherently on $\mathbf{x}$.

\medskip
By implementing these six steps, we successfully construct the global block-encoding $U_\mathbf{K}$ of Eq.\,\eqref{eq:U_K} with the scaling constant $\beta = n_\text{el}\delta$. The complete circuit for a four-element example is shown in Fig.\,\ref{fig:block_encoding}.

\subsection{Choice of QSVT Polynomial}\label{sec:qsvt_poly}
In Section\,\ref{sec:qsvt}, we explained how to use the QSVT algorithm for matrix inversion. Therefore, the QSVT polynomial approximates the reciprocal function ${P_\text{inv}(x)\approx \gamma/x}$ on the interval $x\in[\mu,1]$, and $\mu$ is the smallest singular value that is accurately inverted. 
Following Refs.\,\cite{martyn_grand_2021, gilyen_quantum_2019, childs_quantum_2017}, an odd target function is
\begin{equation}
f_\text{odd}(x)=
\begin{cases}
\dfrac{\mu}{2x}, & |x|\ge \mu,\\[4pt]
\psi_\text{odd}(|x|), & \tfrac{\mu}{2}<|x|<\mu,\\[4pt]
0, & |x|\le \tfrac{\mu}{2},
\end{cases}
\label{eq:f_odd}
\end{equation}
with a smooth interpolation $\psi_\text{odd}:[\mu/2,\mu]\to[0,1/2]$ satisfying $\psi_\text{odd}(\mu/2)=0$ and $\psi_\text{odd}(\mu)=1/2$. Chebyshev techniques yield an $\epsilon\mu/2$-close polynomial approximation to $f_\text{odd}$ of degree \cite{martyn_grand_2021}
\begin{equation}
\deg(P_\text{odd})=\mathcal{O}\!\left(\frac{1}{\mu}\log\frac{1}{\mu\epsilon}\right).
\end{equation}
On $|x|\geq\mu$ this behaves like
\begin{equation}
P_\text{odd}(x)\approx \frac{\mu}{2}\,\frac{1}{x},
\end{equation}
so the effective scale factor is $\gamma_\text{odd}=\mu/2$.

\begin{figure}[h!]
    \centering
    \includegraphics[width=\columnwidth]{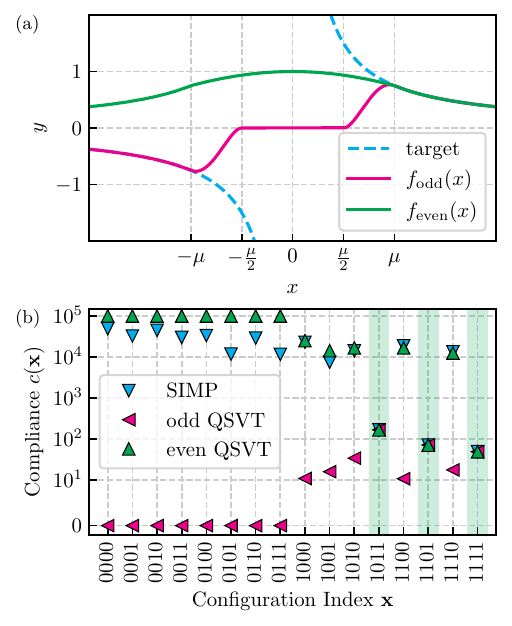}
    \caption{Impact of the QSVT polynomial on compliance computation. (a) Target reciprocal function $\propto 1/x$ compared with its odd and even (polynomial) approximations over the interval $[\mu,1]$. (b) Compliance values for a domain with $n_x=2$, $n_y=2$ obtained using the SIMP approach with $0<x_e\ll1$, and QSVT matrix inversion based on the odd or even approximation function. The green shaded region marks the three feasible configurations. One can observe that the odd QSVT leads to wrongfully small compliances for invalid structures. Consequently, Grover's algorithm would mark all configurations as solutions.}
    \label{fig:compliance_qsvt}
\end{figure}

However, in TO the global stiffness matrix $\mathbf{K}$ can become singular by design if elements are void or supports are insufficient. Then a singular value with $\sigma_i=0$ is possible. Because $f_\text{odd}(0)=0$, loads aligned with such singular vectors are mapped to zero rather than very large displacements, which underestimates compliance for infeasible configurations.
To robustly penalize near-singular directions while keeping $|P_\text{inv}|\leq 1$, we use an even target function that acts as a singular value filter
\begin{equation}
    f_\text{even}(x)=
    \begin{cases}
    \psi_\text{even}(x), & |x|<\mu,\\[6pt]
    \dfrac{y_0\,\mu}{|x|}, & |x|\ge \mu,
    \end{cases}
    \label{eq:even}
\end{equation}
where $\psi_\text{even}(x)$ is a smooth and symmetric function satisfying $\psi_\text{even}(0)=1$ and $\psi_\text{even}(x_1)>\psi_\text{even}(x_2)$ for $0\leq\abs{x_1}\leq\abs{x_2}\leq\mu$, and $y_0=f_\text{even}(\mu)\in(0,1]$. On $|x|\ge\mu$ the polynomial approximation matches the scaled reciprocal,
\begin{equation}
P_\text{even}(x)\approx \frac{y_0\,\mu}{|x|},
\end{equation}
As with the odd case, one can construct a Chebyshev approximation on $[\mu,1]$ with
\begin{equation}
\deg(P_\text{even})=\mathcal{O}\!\left(\frac{1}{\mu}\log\frac{1}{\mu\epsilon}\right).
\label{eq:P_even}
\end{equation}
Fig.\,\ref{fig:compliance_qsvt}(a) shows $f_\text{odd}(x)$ and $f_\text{even}(x)$ scaled so that the function values lie in $[-1, 1]$ and match the reciprocal target function in $[\mu, 1]$. 

In the end, our quantum approach computes the phase $\theta(\mathbf{x})$ from Eq.\,\eqref{eq:phase_cost} associated with
\begin{equation}
    \tilde{c}(\mathbf{x})=\mathbf{f}^\top \mathbf{V}P_\text{inv}(\mathbf{\Sigma})\mathbf{V}^\dagger\mathbf{f},
\end{equation}
where $\tfrac{1}{\beta}\mathbf{K}(\mathbf{x})=\mathbf{V}\mathbf{\Sigma}\mathbf{V}^\dagger$. For feasible structures, $P_\text{inv}(x)\approx \gamma/x$ gives the exact compliance up to a known scaling factor $\tilde{c}(\mathbf{x})\approx \gamma\beta\,c(\mathbf{x})$. For infeasible designs with very small or zero singular values, the even polynomial saturates to 1, yielding large but finite displacements and thus a reasonable objective that mirrors the classical blow-up without numerical instabilities. In the SIMP method, one typically avoids the singular values with $\sigma_i=0$ by setting $0<x_e\ll1$ for void elements. Fig.\,\ref{fig:compliance_qsvt}(b) reports compliance for all configurations of a $2\times2$ MBB beam, where we fixed all horizontal DoFs on the left boundary
and the vertical DoF at the bottom-right corner. All three methods (SIMP with $x_e\in\{0.001,1\}$, odd QSVT with $\mu=10^{-3}$, and even QSVT with $\mu=10^{-3}$ and $y_0=0.01$) agree on the three feasible designs, while the odd QSVT underestimates compliance elsewhere. This would mark all material distributions $\mathbf{x}$ as valid solutions. The even QSVT leads to similarly large compliances on infeasible configurations as the SIMP method, making it a suitable choice for search and filtering.

\subsection{Complexity Analysis}\label{sec:complexity}
We summarize the algorithmic complexities $\mathcal{G}$ of the main subroutines and then combine them into end-to-end expressions. Therefore, we count logical resources only and do not account for error-correction overhead. In the following, we will use equal signs in combination with big O notation for convenience.

Grover's algorithm consists of the initialization $U_\text{init}$ and $r=\mathcal{O}(\sqrt{N/M})$ (cf. Eq.\,\eqref{eq:r}) Grover iterations of $DO$, where $N=\mathcal{O}(2^{n_\text{el}})$ is the size of the search space, and $M$ is the number of marked solutions. Without volume constraints, $U_\text{init}=H^{\otimes n_\text{el}}$ and $\mathcal{G}[U_\text{init}]=n_\text{el}$. With volume constraints, $U_\text{init}$ prepares the Dicke state and $\mathcal{G}[U_\text{init}]=\mathcal{O}(n_\text{el}k)$ \cite{gasieniec_deterministic_2019}. The number of solid elements $k$ depends on how fine our FEM discretization is and thus, we assume $k=\mathcal{O}(n_\text{el})$, which gives $\mathcal{G}[U_\text{init}]=\mathcal{O}(n_\text{el}^2)$. Grover's Diffuser $D$ from Eq.\,\eqref{eq:D} has similar gate complexity $\mathcal{G}[D]=\mathcal{G}[U_\text{init}] + \mathcal{G}[S_0] = \mathcal{O}(n_\text{el}^2)$ since the reflection $S_0$ corresponds to a multi-controlled $Z$ operation with $n_\text{el}$ controls, which can be implemented in $\mathcal{O}(n_\text{el})$ steps \cite{barenco_elementary_1995}. In general, we assume that any unitary $U$ controlled on $n$ qubits can be implemented in $\mathcal{O}(n+\mathcal{G}[U])$ steps by introducing an ancilla qubit \cite{barenco_elementary_1995}. Grover's oracle $O$ (cf. Fig.\,\ref{fig:groversoracle}) has complexity $\mathcal{G}[O] = \mathcal{G}[C] + \mathcal{G}[U_<]$, where $U_<$ is the bitwise comparator to a classical number. If $U_<$ is implemented via a ripple-carry adder, it can be implemented with $\mathcal{O}(n_\texttt{p})$ operations \cite{cuccaro_new_2004}. Here, $n_i, i\in\{\texttt{c, p, l, d, g, h, q, v, z}\}$ is the number of qubits of the respective register. Consequently, the gate complexity of Grover's algorithm is given by
\begin{equation}
\begin{split}
    \mathcal{G}[\text{Grover}] &= \mathcal{G}\left[U_\text{init}\right] + \mathcal{O}(r)(\mathcal{G}[D]+ \mathcal{G}[O])\\
    &= \mathcal{O}\left(\sqrt{N/M}\left(n_\text{el}^2 + n_p + \mathcal{G}[C]\right)\right).
\end{split}
\end{equation}

The operation $C$ computes the compliance $c(\mathbf{x})$ or equivalently the phase $\theta(\mathbf{x})$ from Eq.\,\eqref{eq:phase_cost}. It is a QAE circuit consisting of $n_\texttt{p}$ Hadamards, a single $A$, $\mathcal{O}(2^{n_\texttt{p}})$ controlled Grover operators $G$, and an inverse QFT with $\mathcal{O}(n_\texttt{p})$ operations. The gate complexity of $G$ is dominated by $A$ and is thus given by $\mathcal{G}[G]=\mathcal{G}[A]$. As can be seen in Fig.\,\ref{fig:hadamardtest}, $\mathcal{G}[A] = \mathcal{O}(n_\text{el}+\mathcal{G}[U_{\mathbf{K}^{-1}}])$. Here, we assumed that the state preparation $V_f$ requires $\mathcal{O}(n_\text{el})$ operations \cite{mottonen_transformation_2004}. Accordingly, we write the complexity of $C$ as
\begin{equation}
    \begin{split}
        \mathcal{G}[C] &= \mathcal{O}(\mathcal{G}[H^{\otimes n_\texttt{p}}] + \mathcal{G}[A] + 2^{n_\texttt{p}} \mathcal{G}[G] + \mathcal{G}[\text{QFT}^\dagger])\\
        &=\mathcal{O}(2^{n_\texttt{p}} (n_\text{el}+\mathcal{G}[U_{\mathbf{K}^{-1}}])).
    \end{split}
\end{equation}

The operator $U_{\mathbf{K}^{-1}}$ block-encoding the inverse global stiffness matrix $\mathbf{K}^{-1}$ is the QSVT of $U_\mathbf{K}$ with an even polynomial $P_\text{even}(x)$ approximating $f_\text{even}(x)$ from Eq.\,\eqref{eq:even}. Thus, we have $\deg(P_\text{even})$ calls to $U_\mathbf{K}$ and $\Pi_\phi$. For $n_\text{fixed}$ DoFs, the projector-controlled phase-shift contains $\mathcal{O}(n_\text{fixed})$ multi-controlled $X$ gates (MC$X$s) with up to $n_\texttt{d} + n_\texttt{l} + 3$ controls. Both, the data register $\texttt{d}$ and the $\texttt{l}$ register have $n_\texttt{d}=n_\texttt{l}=\mathcal{O}(\log n_\text{el})$ qubits. Hence, we have
\begin{equation}
    \begin{split}
        \mathcal{G}[U_{\mathbf{K}^{-1}}] &= \deg(P_\text{even}) (\mathcal{G}[U_\mathbf{K}]+\mathcal{G}[\Pi_\phi])\\
        &= \mathcal{O}\left(\tfrac{1}{\mu}\log\tfrac{1}{\mu\epsilon}(\mathcal{G}[U_\mathbf{K}] + n_\text{fixed}\log n_\text{el})\right).
    \end{split}
\end{equation}

The block-encoding operator of the global stiffness matrix $U_{\mathbf{K}}$ requires the following gate operations. The LCU preparation step consists of $\mathcal{O}(n_\texttt{l})$ Hadamards. Moreover, we have $\mathcal{O}(n_\text{el})$ adders $P_{\pm\Delta(e)}$ acting on $n_\texttt{d}$ qubits controlled on $n_\texttt{l}$ qubits. As previously mentioned, we assume ripple-carry adder implementations for estimating algorithmic complexity. This leads to linear complexity in the number of qubits \cite{cuccaro_new_2004}. The conditioning step on $\ket{\mathbf{x}}_\texttt{c}$ introduces $n_\text{el}$ MC$X$s (Multi-Controlled NOT) with $n_\texttt{l}+1$ controls. In addition, we have the permutation operators $P_\text{gap}$ and $P_\text{gap}^\dagger$ consisting of adders on $n_\texttt{d}$ qubits. The block-encoding operator for the element stiffness matrix $U_{\mathbf{K}^\text{el}}$ is implemented in $\mathcal{O}(1)$ steps, $F$ (see Section\,\ref{sec:zero_padding}) corresponds to a MC$X$ with $n_\texttt{d}-3$ controls. Under these considerations, we obtain the complexity
\begin{equation}
    \begin{split}
        \mathcal{G}[U_\mathbf{K}] =& \,\mathcal{G}[H^{\otimes n_\texttt{l}}] + \mathcal{O}(n_\text{el})\,\mathcal{G}[\text{MC}P_{\pm\Delta(e)}]\\ 
        &+ \mathcal{O}(n_\text{el})\,\mathcal{G}[\text{MC}X] +\mathcal{G}[P_\text{gap}] + \mathcal{G}[U_{\mathbf{K}^\text{el}}] + \mathcal{G}[F]\\
        =&\, \mathcal{O}(n_\text{el}\log n_\text{el}).
    \end{split}
\end{equation}
This shows that $U_{\mathbf{K}}$ can efficiently implement $2^{n_\text{el}}$ possible global stiffness matrices $\mathbf{K}(\mathbf{x})$ in only $\mathcal{O}(n_\text{el}\log n_\text{el})$ steps.

Plugging everything together, we get the end-to-end complexity of our fault-tolerant quantum approach to TO
\begin{multline}
\mathcal{G}[\text{Grover}] =
\mathcal{O}\biggl(\sqrt{\tfrac{N}{M}}\Bigl( n_\text{el}^2 + 2^{n_\texttt{p}} \tfrac{1}{\mu}\log\tfrac{1}{\mu\epsilon} (n_\text{el}\log n_\text{el} \\
 + n_\text{fixed}\log n_\text{el})
\Bigr)\biggr).
\label{eq:Grover_preliminary_complexity}
\end{multline}
We can further simplify this expression by expressing the variables $n_\text{fixed}$, $\mu$, and $n_\texttt{p}$ in terms of $n_\text{el}$. The number of fixed DoFs scales like $n_\text{fixed}=\mathcal{O}(n_\text{el})$. The smallest singular value $\mu$ that we need to resolve within QSVT depends on the smallest non-zero eigenvalue within $\mathbf{K}$. In general, we can assume that the smallest non-zero eigenvalue scales like $\mathcal{O}(1/n_\text{el})$ \cite{fried_bounds_1973, lennard_kamenski_study_2014}. In addition, the sum in Eq.\,\eqref{eq:lcusum} gives another scaling with $1/n_\text{el}$. Thus, $\mu=\mathcal{O}(1/n_\text{el}^2)$. Finally, we need to estimate the number of phase qubits $n_\texttt{p}$ storing the phase $\theta(\mathbf{x})$ associated with $c(\mathbf{x})$ as shown in Eq.\,\eqref{eq:phase_cost}. The compliance $c(\mathbf{x})$ converges for fine discretizations with large $n_\text{el}$ and is thus independent of $n_\text{el}$. If we assume a normalized force vector $\mathbf{f}$, the scaling coefficients are $\alpha=1/(\gamma\beta)=\mathcal{O}(n_\text{el})$. This leads to 
\begin{equation}
    \theta(\mathbf{x}) = 1/4 + \mathcal{O}(c(\mathbf{x})/n_\text{el}).
    \label{eq:theta_scaling}
\end{equation}
Hence, $2^{n_\texttt{p}} =\mathcal{O}(n_\text{el})$, and we have the simplified complexities 
\begin{align}
    \mathcal{G}[U_\mathbf{K}] &= \mathcal{O}(n_\text{el}\log n_\text{el}),\label{eq:U_K_complexity}\\
    \mathcal{G}[U_{\mathbf{K}^{-1}}] &= \mathcal{O}(n_\text{el}^3\log n_\text{el}\log(n_\text{el}/\epsilon)),\\
    \mathcal{G}[C] &= \mathcal{O}(n_\text{el}^4\log n_\text{el}\log(n_\text{el}/\epsilon)),\label{eq:compliance_complexity}\\
    \mathcal{G}[\text{Grover}] &=\mathcal{O}\left(\sqrt{N/M}\,n_\text{el}^4\log n_\text{el}\log(n_\text{el}/\epsilon)\right).\label{eq:overall_complexity}
\end{align}

For completeness, we also give the space complexities for each qubit register
\begin{equation}
\begin{split}
    n_\texttt{c} &= n_\text{el},\\
    n_\texttt{p} &= \mathcal{O}(n_\text{el}),\\
    n_\texttt{l} &= \mathcal{O}(\log n_\text{el}),\\
    n_\texttt{d} &= \mathcal{O}(\log n_\text{el}),\\
    n_\texttt{g} &= n_\texttt{h} = n_\texttt{q} = n_\texttt{v} = n_\texttt{z} = 1.
\end{split}
\end{equation}
Thus, the algorithm requires $\mathcal{O}(n_\text{el})$ logical qubits overall, dominated by the phase register.

\section{Numerical Experiments}\label{sec:numerics}
We evaluate our algorithm through numerical simulations with the PennyLane quantum computing framework \cite{bergholm_pennylane_2022}. All experiments are performed by classically simulating the quantum circuits, since the fault-tolerant nature of our approach makes it unsuitable for current NISQ devices. We use material parameters $E=1.0$ and $\nu=0.3$, with horizontal supports along the left edge, a vertical support at the bottom-right corner, and a unit force pointing downwards at the top-left corner, matching the setup of Fig.\,\ref{fig:MBB_example}. We proceed step by step: first verifying compliance computation on a $2\times 2$ domain, then demonstrating Grover's Search over the full binary space, and finally incorporating volume constraints with Dicke initialization on a $3\times 3$ domain.

\subsection{Compliance Computation}\label{sec:exp_compliance_computation}
In the first experiment, we compute the compliance $c(\mathbf{x})$ or, more precisely, the phase $\theta(\mathbf{x})$ defined in Section\,\ref{sec:compliancecomputation}. The corresponding operation $C$ is a QAE circuit in which the Grover operator $G$ contains the QSVT-based matrix inversion $U_{\mathbf{K}^{-1}}$ of the global stiffness matrix $\mathbf{K}$. Because $C$ requires $\mathcal{O}(2^{n_\texttt{p}})$ calls to $U_{\mathbf{K}^{-1}}$, direct simulation quickly becomes intractable. We therefore split the computation into two steps. First, we simulate the QSVT circuit to obtain the (pseudo-)inverse of $\mathbf{K}(\mathbf{x})$ for all material configurations. In a second step, we embed these (pseudo-)inverses as a block-encoding unitary into the Hadamard test (cf. Fig.\,\ref{fig:hadamardtest}) to construct $A$ and $G$. See Appendix\,\ref{sec:details_cc} for further implementation details.

For demonstration, we consider a $2\times2$ domain without a volume constraint, yielding 16 possible configurations. The even QSVT polynomial is obtained by fitting a Chebyshev approximation to Eq.\,\eqref{eq:even} with $\mu=10^{-3}$ and $y_0=0.3$, subject to an absolute error tolerance of $10^{-3}$. The resulting polynomial has degree ${\deg(P_\text{even})=6610}$, which is acceptable for proof-of-concept simulation. Further technical details on the polynomial construction are provided in Appendix\,\ref{sec:poly}. 

Figure\,\ref{fig:compliance_computation}(a) shows the probability distribution of the phase register with $n_\texttt{p}=5$ qubits after applying $C\ket{0}_\texttt{p}\ket{1111}_\texttt{c}$. 
\begin{figure}[h]
    \centering
    \includegraphics{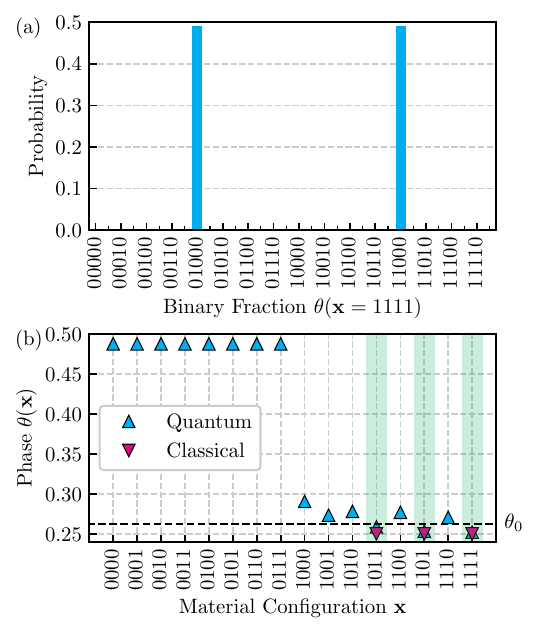}
    \caption{Quantum compliance computation for the $2\times 2$ MBB beam. (a) shows the probability distribution of measuring the phase register $\texttt{p}$ after applying $C\ket{0}_\texttt{p}\ket{1111}_\texttt{c}$. (b) shows $\theta(\mathbf{x})$ for all 16 configurations. Green shading highlights feasible configurations.}
    \label{fig:compliance_computation}
\end{figure}
Two sharp peaks appear at bitstrings ${01000}$ and ${11000}$, corresponding to phases $\pm0.25$ in big-endian notation. The exact value is $\theta(\mathbf{x}=1111)=0.25000826$, validating the simulation. Figure\,\ref{fig:compliance_computation}(b) reports $\theta(\mathbf{x})$ for all 16 configurations. For the three feasible cases $\mathbf{x}\in\{1011,\,1101,\,1111\}$, our quantum circuit results agree with classical FEM values up to the expected polynomial approximation error. Infeasible configurations would exhibit infinite compliance, but our quantum routine maps them to phases corresponding to large, finite values. Setting a threshold $\theta_0=0.263$ cleanly separates feasible from infeasible designs.

For comparison, Fig.\,\ref{fig:compliance_qsvt}(b) presented earlier shows the compliance for the same $2\times2$ domain computed with SIMP regularization, odd QSVT, and even QSVT. There, using smaller values $y_0=0.1$ produces a sharper separation between feasible and infeasible cases at the expense of an even higher polynomial degree as shown in Fig.\,\ref{fig:y0_scaling_analysis}. Additionally, the phase encoding $\theta(\mathbf{x})\in[0.25,0.5)$ compresses compliance values, bringing good and bad designs closer together on the phase scale.

\subsection{Grover's Search}\label{sec:exp_grovers_search}
\begin{figure}[h]
    \centering
    \includegraphics{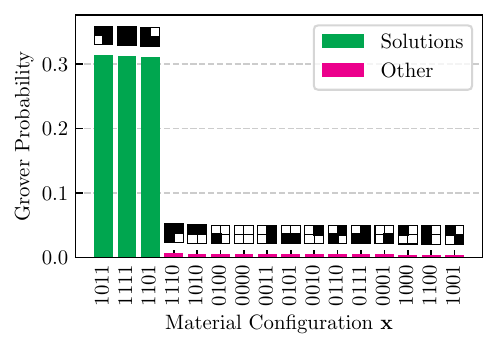}
    \caption{Sorted probability distribution over all 16 configurations after running Grover's algorithm for a $2\times2$ beam. The corresponding structure is shown above each bar, where black elements indicate solid material.}
    \label{fig:grovers_search}
\end{figure}

We next demonstrate Grover's algorithm introduced in Sections\,\ref{sec:groverssearch} and \ref{sec:oracle} on the same $2\times2$ domain. To keep the simulation tractable, we replace the full compliance computation $C$ by a state-preparation routine that directly encodes the QAE outcome for each configuration $\mathbf{x}$ on $n_\texttt{p}=9$ qubits. See Appendix\,\ref{sec:details_gs} for further implementation details. We precompute the phases $\theta(\mathbf{x})$ with QSVT using $\mu=10^{-3}$ and $y_0=0.3$. With a threshold of $\theta_0=0.263$, the oracle marks $M=3$ feasible configurations out of the $N=16$ candidates. After one Grover iteration ($r=1$), their measurement probabilities are significantly amplified, as shown in Fig.\,\ref{fig:grovers_search}. Feasible configurations dominate the distribution, while infeasible ones are strongly suppressed, demonstrating successful amplitude amplification.

\subsection{Volume-Constrained Search}\label{sec:volsearch}
\begin{figure}[h]
    \centering
    \includegraphics{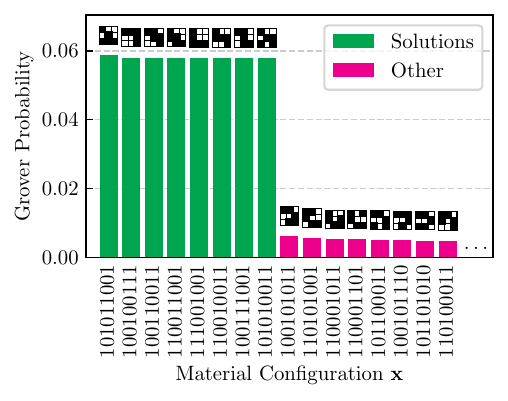}
    \caption{Sorted probability distribution of the 16 most probable configurations for volume-constrained Grover's search on the $3\times3$ beam with $V_0=5/9$ filled elements enforced by Dicke initialization. The corresponding structure is shown above each bar, where black elements indicate solid material.}
    \label{fig:grovers_search_volume_constrained}
\end{figure}

Finally, we extend the FEM discretization to a $3\times3$ domain while enforcing a fixed material volume of $V_0=5/9$ elements via Dicke state initialization ${U_\text{init}\ket{0}_\texttt{c}=|D^{(9)}_5\rangle_\texttt{c}}$. This reduces the search space from $2^9=512$ to $N=\binom{9}{5}=126$ configurations. As before, we replace $C$ by a state-preparation routine encoding the QAE outcome on $n_\texttt{p}=9$ qubits. Since the domain is larger with $n_\text{el}=9$ elements, we choose a smaller $\mu=10^{-5}$ (with $y_0=0.3$) for QSVT precomputation. After $r=2$ Grover iterations with threshold $\theta_0=0.251$, the algorithm identifies $M=8$ feasible configurations. Fig.\,\ref{fig:grovers_search_volume_constrained} shows the resulting probability distribution, which correctly highlights all eight solutions.

The number of phase qubits $n_\texttt{p}=9$ is sufficient to distinguish between feasible and infeasible configurations, with phases up to $\theta=0.2506$ and above $\theta=0.2553$, respectively. However, resolving the optimal configuration among the eight feasible designs would require distinguishing $\Delta\theta\approx5\times10^{-6}$, which would in turn demands at least $n_\texttt{p}=18$ qubits. From Eq.\,\eqref{eq:theta_scaling}, we know that $\theta(\mathbf{x})$ approaches $0.25$ as the number of elements $n_\text{el}$ increases, explaining the high resolution required. 

This limitation can be reframed as a feature. If our goal is only to identify feasible solutions under the volume constraint (rather than the single optimal solution), we can intentionally use fewer phase qubits. This effectively rounds all feasible phases to $\theta=0.25$, as in our experiment above where $\theta=0.2506\approx0.25$ is rounded. A side effect is that the QAE probability distribution exhibits smoother amplitude accumulation instead of sharp peaks, which introduces noise into the Grover oracle. This explains why the separation between solution and non-solution probabilities is less pronounced here than in the unconstrained experiment.

\section{Discussion}\label{sec:discussion}
While our numerical experiments showcase the functionality of the proposed algorithm, our complexity analysis illustrates where it has an advantage compared to classical approaches. The compliance computation $C$ consists of the QSVT-based matrix inversion within the QAE circuit. 
It can in principle evaluate $N=\mathcal{O}(2^{n_\text{el}})$ candidate designs in polynomial time in $n_\text{el}$, see Eq.\,\eqref{eq:compliance_complexity}. For a single linear system, this offers no advantage over state-of-the-art classical linear equation system solvers, such as Gaussian elimination with $\mathcal{O}(n_\text{el}^3)$ or the conjugate gradient method with $\mathcal{O}(n_\text{el}\sqrt{\kappa})$ \cite{golub_matrix_2013} in optimal cases, where $\kappa=\mathcal{O}(n_\text{el})$ \cite{fried_bounds_1973, lennard_kamenski_study_2014} is the condition number. However, since we never want to read out all displacement fields $\mathbf{u}(\mathbf{x})=\mathbf{K}^{-1}(\mathbf{x})\mathbf{f}$, we can avoid the exponential output overhead \cite{aaronson_read_2015}. Instead we evaluate exponentially many structures in coherent superposition using only $\mathcal{O}(n_\text{el}^4\log n_\text{el}\log(n_\text{el}/\epsilon))$ quantum gates. On top of $C$, Grover's algorithm finds an optimal structure $\mathbf{x}$ satisfying Eq.\,\eqref{eq:satisfiability_problem} in quadratically less steps (cf. Eq.\,\eqref{eq:overall_complexity}) compared to a classical unstructured search algorithm. This poses a significant advantage over classical methods when one seeks not only the global optimum but all feasible solutions to a problem. More than one feasible solution might be desired because of latent requirements that cannot be considered in the optimization formulation. For example constraints from the chosen manufacturing procedure or human-product interaction. In the quantum setting, additional solutions can be found simply by repeated execution and sampling from the final quantum state. In contrast, classical gradient-based solvers typically require repeated experimentation with learning rates, penalty terms, and initialization strategies to find other solutions.

While the theoretical quadratic advantage seems promising, it might not be enough to be translated to an actual practical speedup \cite{babbush_focus_2021}. Moreover, relaxation methods such as SIMP \cite{sigmund_99_2001, andreassen_efficient_2011} avoid exponential runtime by using gradient-based optimization. Although these methods may not find the global optimum, in practice a near-optimal result is often sufficient. Another practical limitation arises from the scaling $\mu=\mathcal{O}(1/n_\text{el}^2)$, which originates from the condition number $\kappa$ and an additional factor of $1/n_\text{el}$ from the LCU in Eq.\,\eqref{eq:lcusum}. The determination of the QSVT phases ${\phi_d}$ for small $\mu$, and thus for high-degree QSVT polynomials, quickly becomes a non-negligible computational task itself \cite{dong_efficient_2021, lin_mathematical_2025, ni_fast_2024}. Moreover, the runtime of QSVT, and consequently of our quantum topology algorithm, scales superlinearly with $1/\mu$ (cf. Eq.\,\eqref{eq:Grover_preliminary_complexity}).

One way to increase $\mu$ for larger $n_\text{el}$ would be to introduce quantum-compatible preconditioners \cite{clader_preconditioned_2013, deiml_quantum_2024, lapworth_preconditioned_2025}. In the best case, this would have insignificant computational overhead and make $\kappa$ independent of $n_\text{el}$, thereby reducing the scaling of $\mu$ by a factor of $1/n_\text{el}$. Another potential improvement could come from optimizing our block-encoding algorithm for $U_\mathbf{K}$ to mitigate the additional $1/n_\text{el}$ factor in the LCU in Eq.\,\eqref{eq:lcusum}. However, it remains unclear how such an optimization could be achieved, as the current block-encoding is already highly efficient. In comparison to a conventional LCU block-encoding of a single stiffness matrix $\mathbf{K}$ with $\mathcal{O}(n_\text{el}^2)$ Pauli strings, our approach requires only $\mathcal{O}(n_\text{el}\log n_\text{el})$ gates. This does not even change when preparing all $2^{n_\text{el}}$ stiffness matrices $\mathbf{K}(\mathbf{x})$ in superposition.

Further improvements of our algorithm could be to reduce the QSVT circuit depth by implementing generalized QSVT \cite{motlagh_generalized_2024}, which extends projector-controlled phase-shifts beyond the $Z$ axis. While we are using the nature of QSVT to handle singular matrices, one might also use other quantum linear system solvers with reduced runtime complexity \cite{costa_optimal_2022, liu_improved_2025, dalzell_shortcut_2024}. Another possible extension would be to evaluate the structural performance of the system through frequency-domain response functions, as recently demonstrated by Danz et al. \cite{danz_calculating_2025}. Their method computes the dynamic response of coupled oscillator systems directly from the FEM stiffness and mass matrices and could complement the present workflow by enabling dynamic or frequency-dependent topology optimization. Also their block-encoding with matrix access oracles \cite{camps_explicit_2024} could represent an attractive alternative to our block-encoding techniques. On the optimization side, Grover's algorithm could be replaced by alternative quantum optimization methods, such as QAOA \cite{farhi_quantum_2014, stein_exponential_2023, holscher_quantum_2025, adler_scaling_2025} or decoded quantum interferometry \cite{jordan_optimization_2024, sabater_towards_2025}. In general, Bennett et al. \cite{bennett_strengths_1997} have shown that Grover's algorithm is asymptotically optimal for unstructured search. Nevertheless, since we can prepare the entire search space in polynomial time, Grover's algorithm becomes the bottleneck of our TO approach. Structured search algorithms that exploit the problem's inherent structure could potentially overcome this limitation.

\section{Conclusion}\label{sec:outlook}

We have presented a fault-tolerant quantum algorithm for TO and demonstrated its feasibility on the 2D MBB beam problem. Our approach combines block-encoding of FEM stiffness matrices, QSVT-based matrix inversion, compliance computation through a Hadamard test and QAE, and Grover's search with Dicke state initialization to enforce volume constraints. By integrating these components, we establish an end-to-end demonstration of how TO can be reformulated and addressed within a coherent quantum workflow. In our detailed complexity analysis, we show that our algorithm maintains the quadratic Grover speedup for unstructured search.

Our numerical experiments validate each subroutine. Compliance computation via QSVT and QAE reproduces classical FEM results within polynomial approximation error, while Grover's search amplifies feasible, low-compliance designs in both unconstrained and volume-constrained settings. These proof-of-concept simulations confirm that the algorithm is logically consistent and illustrate the potential of quantum superposition to evaluate many candidate designs at once. At the same time, they highlight technical challenges, notably the growth of polynomial degree with system size and the high phase resolution required to distinguish near-optimal solutions.

Although our present study is limited to small-scale examples simulated on classical hardware, it establishes a blueprint for fault-tolerant quantum TO. As quantum hardware progresses toward large-scale fault-tolerance, the principles demonstrated here may enable practical advantages for engineering design tasks where the combinatorial explosion of candidate structures is the primary computational bottleneck.

\section*{Data and code availability}
All code and Jupyter notebooks to reproduce the results and figures presented in this work are available at
\href{https://github.com/Emerging-Tech-MUC/QTO}{github.com/Emerging-Tech-MUC/QTO}.

\begin{acknowledgments}
We acknowledge the support of the BMW Group. L. Karch is partly funded by the German Ministry for Economic Affairs and Energy (BMWE) in the project QCHALLenge under Grant 01MQ22008D.
O. Ahrend acknowledges support by the Interdisciplinary Doctoral Program in Quantum Systems Integration funded by the BMW Group. 
\end{acknowledgments}

\appendix

\section{Derivation of Element Stiffness Matrix}\label{appendix:derivationstiffness}
In the following, we provide the exact form of the element stiffness matrix $\mathbf{K^\text{el}}$ for a square,
bilinear, four-noded quadrilateral element. Due to the element's simple geometry, $\mathbf{K}^\text{el}$ is determined by eight unique parameters
\begin{equation}
\resizebox{\columnwidth}{!}{$
\begin{array}{llll}
k_1 = \frac{1}{2}-\frac{\nu}{6}, & k_2 = -\frac{1}{8}-\frac{\nu}{8}, & k_3 = \frac{\nu}{6}, & k_4 = -\frac{1}{8}+\frac{3\nu}{8}, \\
k_5 = -\frac{1}{4}-\frac{\nu}{12}, & k_6 =\frac{1}{8}-\frac{3\nu}{8}, & k_7 = -\frac{1}{4}+\frac{\nu}{12}, & k_8 = \frac{1}{8}+\frac{\nu}{8}
\end{array}
$}
\end{equation}
yielding $\mathbf{K}^\text{el} = \frac{E}{1-\nu^2}\mathbf{K}_0$ with
\begin{equation}
    \mathbf{K}_0=\begin{pmatrix}
        k_{1}&k_{2}&k_{3}&k_{4}&k_{5}&k_{6}&k_{7}&k_{8}\\
        k_{2}&k_{1}&k_{6}&k_{5}&k_{4}&k_{3}&k_{8}&k_{7}\\
        k_{3}&k_{6}&k_{1}&k_{8}&k_{7}&k_{2}&k_{5}&k_{4}\\
        k_{4}&k_{5}&k_{8}&k_{1}&k_{2}&k_{7}&k_{6}&k_{3}\\
        k_{5}&k_{4}&k_{7}&k_{2}&k_{1}&k_{8}&k_{3}&k_{6}\\
        k_{6}&k_{3}&k_{2}&k_{7}&k_{8}&k_{1}&k_{4}&k_{5}\\
        k_{7}&k_{8}&k_{5}&k_{6}&k_{3}&k_{4}&k_{1}&k_{2}\\
        k_{8}&k_{7}&k_{4}&k_{3}&k_{6}&k_{5}&k_{2}&k_{1}
    \end{pmatrix}.
    \label{eq:Kel}
\end{equation}
Here, Young's modulus $E$ and Poisson's ratio $\nu$ are material properties. 

\begin{figure}[h]
    \centering
    \includegraphics{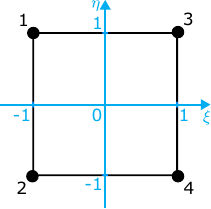}
    \caption{Square, four-noded element defined in the isoparametric coordinate system.}
    \label{fig:quadrilateral_element}
\end{figure}

Next, we outline the derivation of $\mathbf{K}^\text{el}$ following standard FEM analysis.
Fig.\,\ref{fig:quadrilateral_element} shows the element in the
isoparametric coordinate system $(\xi,\eta)\in[-1,1]^2$. 

\noindent The four corner nodes are
\begin{equation}
\begin{array}{lll}
    (1) \text{ top-left:} & \xi=-1, & \eta=1, \\
    (2) \text{ bottom-left:} & \xi=-1, & \eta=-1, \\
    (3) \text{ top-right:} & \xi=1, & \eta=1, \\
    (4) \text{ bottom-right:} & \xi=1, & \eta=-1.
\end{array}
\end{equation}

For a physical element of side length $l$, we have the mapping
\begin{equation}
    x=\frac{l}{2}(\xi+1) \quad\text{and}\quad y=\frac{l}{2}(\eta+1).
\end{equation}
The Jacobi determinant of this transformation is $\abs{\mathbf{J}}=l^2/4$. Using $(\xi,\eta)$ lets one write
a single element formulation that applies to any square or rectangular element. In the end, the factor $\abs{\mathbf{J}}$ rescales the stiffness matrix to its physical dimensions.

Following typical finite element analysis, we define so-called bilinear shape functions for each node as
\begin{equation}
    \begin{array}{ll}
        N_1=\frac{1}{4}(1-\xi)(1+\eta), & N_2=\frac{1}{4}(1-\xi)(1-\eta), \\
        N_3=\frac{1}{4}(1+\xi)(1+\eta), & N_4=\frac{1}{4}(1+\xi)(1-\eta).
    \end{array}
\end{equation}
Each $N_i$ equals one at its own node and zero at the others, so the combination $\sum_i N_i (\,u_{2i-1},u_{2i})^\top$ provides a bilinear interpolation of displacement inside the element.

Next, we calculate the $3\times8$ strain-displacement matrix $\mathbf{B}$ that converts nodal displacements $\mathbf{u}^\text{el}$ to in-plane strains $\mathbf{\varepsilon}=(\varepsilon_{xx},\varepsilon_{yy},\gamma_{xy})^\top$. Using the spatial derivatives
\begin{equation}
    \frac{\partial N_i}{\partial x} = \frac{2}{l}\frac{\partial N_i}{\partial\xi} \quad\text{and}\quad \frac{\partial N_i}{\partial y} = \frac{2}{l}\frac{\partial N_i}{\partial\eta}
\end{equation}
the strain-displacement matrix is given by
\begin{equation}
    \mathbf{B}(\xi, \eta) = \begin{pmatrix}
        \frac{\partial N_1}{\partial x} & 0 & \frac{\partial N_2}{\partial x} & 0 & \frac{\partial N_3}{\partial x} & 0 & \frac{\partial N_4}{\partial x} & 0 \\
        0 & \frac{\partial N_1}{\partial y} & 0 & \frac{\partial N_2}{\partial y} & 0 & \frac{\partial N_3}{\partial y} & 0 & \frac{\partial N_4}{\partial y}\\
        \frac{\partial N_1}{\partial y} & \frac{\partial N_1}{\partial x} & \frac{\partial N_2}{\partial y} & \frac{\partial N_2}{\partial x} & \frac{\partial N_3}{\partial y} & \frac{\partial N_3}{\partial x} & \frac{\partial N_4}{\partial y} & \frac{\partial N_4}{\partial x} 
    \end{pmatrix}.
\end{equation}

For an isotropic, linear-elastic material under plane-stress
conditions, Hooke's law reads
\(\mathbf\sigma=\mathbf D\,\mathbf\varepsilon\) with the constitutive matrix
\begin{equation}
    \mathbf{D} = \frac{E}{1-\nu^2}\begin{pmatrix}
        1 & \nu & 0\\
        \nu & 1 & 0\\
        0 & 0 & \frac{1-\nu}{2}
    \end{pmatrix},
\end{equation}
where Young's modulus $E$ and Poisson's ratio $\nu$ are material parameters.

Finally, the element stiffness matrix is given by
\begin{equation}
    \mathbf{K}^\text{el} = \int_{-1}^{1}\int_{-1}^{1}\mathbf{B}^\top\mathbf{D}\mathbf{B}\abs{\mathbf{J}}d\xi d\eta
\end{equation}
leading to the definition in Eq.\,\eqref{eq:Kel}. Notably, $\mathbf{K}^\text{el}$ is independent of $l$ since the terms in $\mathbf{B}$ and $\abs{\mathbf{J}}$ cancel out.

\section{Practical Construction of the QSVT Polynomial}\label{sec:poly}
Constructing the QSVT polynomial is an essential step in our algorithm. Ref.\,\cite{martyn_grand_2021} describes how one can analytically build an approximation of $f_\text{odd}(x)$ in Eq.\,\eqref{eq:f_odd} by combining known polynomial approximations of elementary functions such as rectangular step functions. This analytical approach has the important advantage that it allows a derivation of the asymptotic behavior of the resulting polynomial. However, in practice such constructions typically lead to very high polynomial degrees. Equivalent approximations of much lower degree can often be obtained numerically. For this reason, we demonstrate the numerical polynomial approximation of $f_\text{even}(x)$ in Eq.\,\eqref{eq:even} in the following, rather than relying on an explicit analytical construction.

While the theoretical form in Eq.\,\eqref{eq:even} specifies the general structure, the practical implementation requires a concrete choice for the smooth interpolation $\psi_{\text{even}}(x)$ in the region $|x| < \mu$. We use the cosine interpolation
\begin{equation}
\psi_{\text{even}}(x) = \cos\left(\tfrac{\arccos(y_0)}{\mu} \, x\right), \quad |x| < \mu,
\end{equation}
which ensures a smooth connection at $x = \pm\mu$ with $f_{\text{even}}(\pm\mu) = y_0$. The complete target function is then
\begin{equation}
f_{\text{even}}(x) =
\begin{cases}
\cos\left(\tfrac{\arccos(y_0)}{\mu} \, x\right), & |x| < \mu, \\[6pt]
\tfrac{y_0 \mu}{|x|}, & |x| \geq \mu,
\end{cases}
\label{eq:f_even_practical}
\end{equation}

We approximate $f_{\text{even}}(x)$ using Chebyshev polynomials. Exploiting the even symmetry of the target function reduces the effective degree of the fitted polynomial by half, improving numerical stability and reducing computation time. Instead of approximating $f_{\text{even}}(x)$ directly on $[-1,1]$, we approximate the auxiliary function $g(t) = f_{\text{even}}(\sqrt{t})$ on $t \in [0,1]$. If
\begin{equation}
P(t) = \sum_{k=0}^n c_k T_k(t)
\end{equation}
is a Chebyshev polynomial approximation of $g(t)$, then
\begin{equation}
Q(x) = P(x^2) = \sum_{k=0}^n c_k T_k(x^2)
\end{equation}
automatically satisfies the symmetry $Q(-x) = Q(x)$ and approximates $f_{\text{even}}(x)$ on $[-1,1]$. In practice, this corresponds to placing the coefficients $c_k$ at even positions in the Chebyshev expansion,
\begin{equation}
Q(x) = \sum_{k=0}^n c_k T_{2k}(x) = c_0 T_0(x) + c_1 T_2(x) + c_2 T_4(x) + \cdots.
\end{equation}

\begin{figure}
    \centering
    \includegraphics{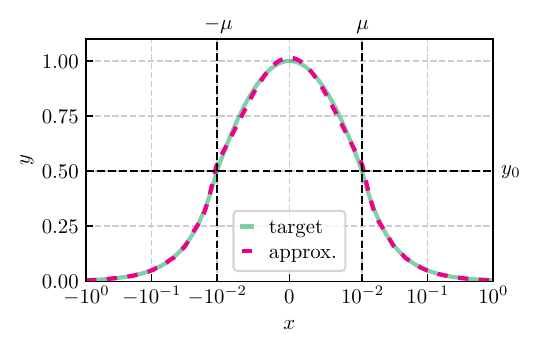}
    \caption{Chebyshev polynomial approximation (degree $382$) of $f_{\text{even}}(x)$ for $\mu = 0.01$, $y_0 = 0.5$, and $\varepsilon = 10^{-3}$. The horizontal axis is shown on a linear scale in the region ${x \in [-\mu,\mu]}$.}
    \label{fig:poly_approx}
\end{figure}

We implemented this Chebyshev approximation procedure using the Python library \texttt{chebpy} \cite{richardson_chebpy_2016}. Fig.\,\ref{fig:poly_approx} shows the even target function for $\mu = 0.01$, $y_0 = 0.5$, and a polynomial approximation with an absolute error tolerance of $\varepsilon = 10^{-3}$. The resulting polynomial has degree $382$ and overlaps well with the target function.

\begin{figure}
    \centering
    \includegraphics{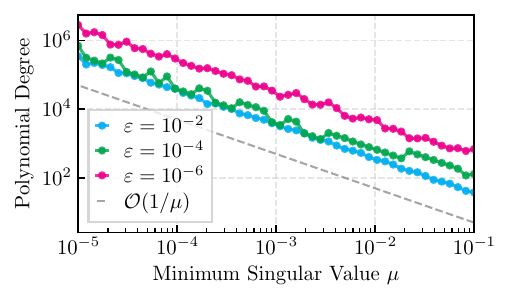}
    \caption{Scaling of the polynomial degree with $\mu$ for different error tolerances $\varepsilon$. The observed behavior confirms the $\mathcal{O}(1/\mu)$ stated in Eq.\,\eqref{eq:P_even}.}
    \label{fig:mu_scaling_analysis}
\end{figure}

Using this numerical construction, we study how the required polynomial degree scales with $\mu$, $\varepsilon$, and $y_0$. The parameter $\mu$, representing the smallest singular value inverted accurately, has a significant impact on runtime complexity. Fig.\,\ref{fig:mu_scaling_analysis} shows the degree of the polynomial against $\mu$ for different $\varepsilon$, confirming the expected reciprocal scaling $\mathcal{O}(1/\mu)$ assumed in Eq.\,\eqref{eq:P_even}.

\begin{figure}
    \centering
    \includegraphics{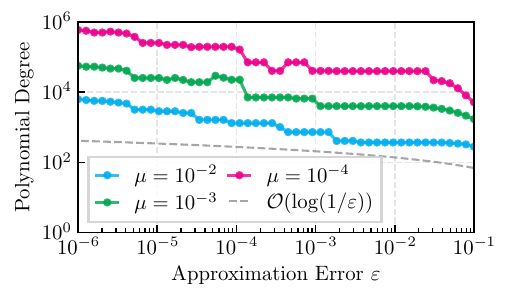}
    \caption{Scaling of the polynomial degree with approximation error $\varepsilon$. The results agree with the expected $\mathcal{O}(\log(1/\varepsilon))$ scaling.}
    \label{fig:eps_scaling_analysis}
\end{figure}

Similarly, Fig.\,\ref{fig:eps_scaling_analysis} shows the polynomial degree as a function of $\varepsilon$, agreeing with the expected $\mathcal{O}(\log(1/\varepsilon))$ scaling. Note that Eq.\,\eqref{eq:P_even} characterizes the case of relative error $\epsilon = \mathcal{O}(\varepsilon / \mu)$, whereas $\varepsilon$ is the absolute error.

\begin{figure}[H]
    \centering
    \includegraphics{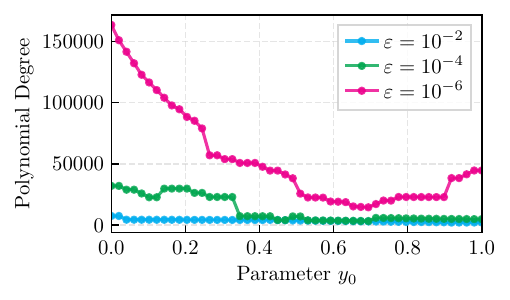}
    \caption{Influence of $y_0$ on the required polynomial degree. Although no clear scaling law is visible, the polynomial degree typically increases with decreasing $y_0$.}
    \label{fig:y0_scaling_analysis}
\end{figure}

Finally, we analyze the influence of $y_0$ in Fig.\,\ref{fig:y0_scaling_analysis}. No clear scaling law emerges, but since $y_0$ can be chosen independently of $\mu$ and $\varepsilon$, we typically fix it to a moderate constant, e.g. $y_0 = 0.3$. The parameter $y_0$ controls the distance in $\theta(\mathbf{x})$ between feasible and infeasible solutions. Thus, in practice we aim to set $y_0$ as small as possible while still keeping the polynomial degree manageable.

Consequently, our numerical construction of $f_{\text{even}}(x)$ using Chebyshev polynomials provides a practical and efficient way to obtain QSVT polynomials of moderate degree while retaining the expected asymptotic scaling with $\mu$ and $\varepsilon$. This validates the theoretical complexity estimates given in Sec.\,\ref{sec:qsvt}, and confirms that the polynomial construction step does not introduce unexpected overhead beyond the known $\mathcal{O}(1/\mu)$ and $\mathcal{O}(\log(1/\varepsilon))$ behavior. 

\section{Implementation Details of Our Numerical Experiments}
The end-to-end algorithm presented here yields a deep quantum circuit acting on many qubits. Its full execution would require fault-tolerant quantum computers, and its simulation quickly becomes intractable. Nevertheless, we presented simulation results of our algorithm in Section\,\ref{sec:numerics}. To make these simulations feasible, we replaced certain costly quantum operations by hardcoded unitaries. While this makes simulations of small systems (with few elements in the domain discretization) tractable, this approach of course does not work for executions on real machines and is not scalable for large simulations, as replacing these unitaries can lead to exponential overhead. In the following, we provide the exact details on how we simplified the simulations for our numerical experiments.

\subsection{Details on Compliance Computation}\label{sec:details_cc}
In Section\,\ref{sec:exp_compliance_computation}, we split the compliance computation $C$ in two steps. First, we use QSVT to calculate the (pseudo-)inverses $\mathbf{K}^{-1}(\mathbf{x})$ for each material configuration $\mathbf{x}$. Therefore, we use \texttt{chebpy} \cite{richardson_chebpy_2016} to fit a Chebyshev polynomial to our even target function in Eq.\,\eqref{eq:f_even_practical} with $\mu=10^{-3}$ and an absolute error $\varepsilon=10^{-3}$. This leads to a polynomial with degree $6610$. We then convert the Chebyshev coefficients to the set of phase angles $\{\phi_k\}$ required for the projector-controlled phase-shifts $\Pi_{\phi_k}$ using the Python library \texttt{pyqsp} \cite{martyn_grand_2021}. This allows us to calculate $U_{\mathbf{K}^{-1}}$ from Eq.\,\eqref{eq:U_K_inverse} and then to extract $\mathbf{K}^{-1}$.

In a second step, we manually block-encode $\mathbf{K}^{-1}$ in a unitary operator as shown by Eq.\,\eqref{eq:U_Kel}. Thereby, we get a unitary $U_{\mathbf{K}^{-1}}$ which replaces the deep QSVT circuit within $A$ (cf. Fig.\,\ref{fig:hadamardtest}) to compute the entire operation $C$.

\subsection{Details on Grover's Search}\label{sec:details_gs}
In Section\,\ref{sec:exp_grovers_search}, we demonstrate Grover's Search for a $2\times2$ domain. There, we replace the compliance computation $C$ by several multi-controlled state preparation routines. 
As shown in Fig.\,\ref{fig:compliance_computation}, $C$ is a QAE circuit which estimates the absolute value of an amplitude $a$ of some state
\begin{equation}
    \ket{\phi} = a\ket{\alpha}_\texttt{d} + \sqrt{1-a^2}\ket{\alpha^\perp}_\texttt{d}.
\end{equation}

If we know $\abs{a}$, we also know the analytical expression of state after QAE
\begin{align}
\begin{split}
    \mathrm{QAE}\ket{0}_\texttt{p}\ket{\phi}_\texttt{d} &= -\frac{i}{\sqrt{2}}\sum_{j=0}^{2^{n_\texttt{p}}-1}e^{-i\pi\theta}\alpha_+(j)\ket{j}_\texttt{p}\ket{\psi_+}_\texttt{d}\\
    &\quad-e^{i\pi\theta}\alpha_-(j)\ket{j}_\texttt{p}\ket{\psi_-}_\texttt{d},
\end{split}
\label{eq:QAE_analytical}
\end{align}
where the phase $\theta$ is given by
\begin{equation}
    \theta = \pm\tfrac{1}{\pi}\arcsin(\abs{a}),
\end{equation}
the amplitudes $\alpha_\pm(j)$ are
\begin{equation}
    \alpha_\pm(j) = \frac{1}{2^{n_\texttt{p}}}\sum_{l=0}^{2^{n_\texttt{p}}-1}e^{-i2\pi l(2^{-{n_\texttt{p}}}j\mp\theta)},
\end{equation}
and 
\begin{equation}
    \ket{\psi_\pm} = \frac{1}{\sqrt{2}}(\ket{\alpha}\pm i\ket{\alpha^\perp})
\end{equation}
are the eigenstates of the Grover operator $G$ shown in Fig.\,\ref{fig:groveroperator} in the subspace spanned by $\ket{\alpha}$ and $\ket{\alpha^\perp}$.
In our TO case, $\abs{h_0(\mathbf{x})}$ in Eq.\,\eqref{eq:h0} is the absolute amplitude $\abs{a}$ that we would like to estimate. We classically calculate 
\begin{equation}
    \abs{h_0(\mathbf{x})} = \sqrt{\tfrac{1}{2}+\tfrac{1}{2\alpha}\mathbf{f}^\top\underbrace{\tfrac{\gamma}{\beta}\mathbf{K}^{-1}(\mathbf{x})}_{=\mathbf{V}f_\text{even}(\mathbf{\Sigma})\mathbf{V}^\dagger}\mathbf{f}}
\end{equation}
for every configuration $\mathbf{x}$.
Thereby, we also replace the deep QSVT circuit by its expected outcome $\mathbf{V}f_\text{even}(\mathbf{\Sigma})\mathbf{V}^\dagger$. Note that we directly use the target function $f_\text{even}(\cdot)$ from Eq.\,\eqref{eq:f_even_practical} instead of fitting a Chebyshev polynomial to it. We use all values $\abs{h_0(\mathbf{x})}$ to calculate the QAE amplitudes according to Eq.\,\eqref{eq:QAE_analytical} and prepare the quantum state accordingly controlled on the respective configuration $\ket{\mathbf{x}}_\texttt{c}$. This allows us to embed these multi-controlled state preparations in Grover's oracle (cf. Fig.\,\ref{fig:groversoracle}) replacing $C$.

\subsection{Details on Volume-Constrained Search}\label{sec:details_vcs}
In Section\,\ref{sec:volsearch}, we demonstrate Grover's search for a $3\times3$ domain with an additional volume constraint. Grover's oracle is simulated as in the previous unconstrained $2\times2$ case. However, the operation $U_\text{init}$ in Grover's algorithm needs to prepare the right Dicke state, whereas it consisted of simple Hadamards before. Instead of the efficient routine of Ref.\,\cite{gasieniec_deterministic_2019}, we employ a standard state preparation operator provided by PennyLane. This suffices for our purpose of demonstrating the algorithm. 
\bibliography{references_sync}

\end{document}